\documentclass{article} 
\usepackage{aaai2026}  
\usepackage{amsmath}
\usepackage[colorlinks=true, linkcolor=blue]{hyperref} 
\usepackage{cleveref}
 
\usepackage{times}  
\usepackage{helvet}  
\usepackage{courier}  
\usepackage{graphicx} 
\usepackage{natbib}  
\usepackage{caption} 
\usepackage{booktabs}
\usepackage{multirow}
\usepackage{makecell}
\usepackage{stackengine} 
\usepackage{algorithm}
\usepackage{algorithmic}

\usepackage{newfloat}
\usepackage{listings}
\DeclareCaptionStyle{ruled}{labelfont=normalfont,labelsep=colon,strut=off} 
\floatstyle{ruled}
\newfloat{listing}{tb}{lst}{}
\floatname{listing}{Listing}
\title{AudioCodecBench: A Comprehensive Benchmark for Audio Codec Evaluation}
\author {
    Lu Wang\textsuperscript{\rm 1},
    Hao Chen\textsuperscript{\rm 1}\thanks{These authors are core and equal contributors},
    Siyu Wu\textsuperscript{\rm 1}\footnotemark[1],
    Zhiyue Wu\textsuperscript{\rm 1}\footnotemark[1]\thanks{Project lead},
    Hao Zhou\textsuperscript{\rm 2},
    Chengfeng Zhang\textsuperscript{\rm 3},
    Ting Wang\textsuperscript{\rm 4},
    Haodi Zhang\textsuperscript{\rm 1} 
}

\affiliations {
    \textsuperscript{\rm 1} Shenzhen University   
    \textsuperscript{\rm 2} Nankai University 
    \textsuperscript{\rm 3} Zhejiang University 
    \textsuperscript{\rm 4} East China Normal University \\
    \{wanglu, hdzhang\}@szu.edu.cn, twang@sei.ecnu.edu.cn
}

\usepackage{bibentry}

\begin{document}

\maketitle

\begin{abstract}
Multimodal Large Language Models (MLLMs) have been widely applied in speech and music. This tendency has led to a focus on audio tokenization for Large Models (LMs). Unlike semantic-only text tokens, audio tokens must both capture global semantic content and preserve fine-grained acoustic details. Moreover, they provide a discrete method for speech and music that can be effectively integrated into MLLMs. However, existing research is unsuitable in the definitions of semantic tokens and acoustic tokens. In addition, the evaluation of different codecs typically concentrates on specific domains or tasks, such as reconstruction or Automatic Speech Recognition (ASR) task, which prevents fair and comprehensive comparisons. To address these problems, this paper provides suitable definitions for semantic and acoustic tokens and introduces a systematic evaluation framework. This framework allows for a comprehensive assessment of codecs' capabilities which evaluate across four dimensions: audio reconstruction metric, codebook index (ID) stability, decoder-only transformer perplexity, and performance on downstream probe tasks. Our results show the correctness of the provided suitable definitions and the correlation among reconstruction metrics, codebook ID stability, downstream probe tasks and perplexity.

\end{abstract}

\begin{links}
     \textbf{Code}: \href{https://github.com/wuzhiyue111/Codec-Evaluation}{https://github.com/wuzhiyue111/Codec-Evaluation} \\
     \textbf{Dataset}: \href{https://huggingface.co/datasets/LeBeGut/AudioCodecBench}{https://huggingface.co/datasets/LeBeGut/Audio\\CodecBench}
\end{links}

\section{Introduction}
Discrete audio tokens have received attention for their potential to bridge the domains of text and audio, playing an important role in the development of Multimodal Large Language Models (MLLMs)~\nocite{i1-1, i-2}\citep{i1-1, i-2}. The process of generating discrete token is compressing the original waveform into a finite set of vectors. However, MLLMs focus more on semantic in the text domain, but need to focus on both semantic and acoustic in the audio domain, resulting in a modality gap between text and audio.  Semantic tokens are often obtained through the quantization hidden states from Self-supervised Learning (SSL) models. These tokens fixed patterns in the same semantic informations so that the fixed patterns are easier to be modeled by downstream tasks~\nocite{i3}\citep{i3}. Acoustic tokens are often obtained by training the neural audio codec (Codecs) in an end-to-end manner with the goal of high-fidelity reconstruction. These tokens focus more on the absolute distance between audio sampling points. This absolute distance definitely contains semantic, but this part of the semantic is difficult to be modeled in downstream tasks and is more suitable for reconstruction~\nocite{i12, i6}\citep{i12, i6}. 

\begin{figure}[t]
    \centering
    \includegraphics[width=0.9\linewidth]{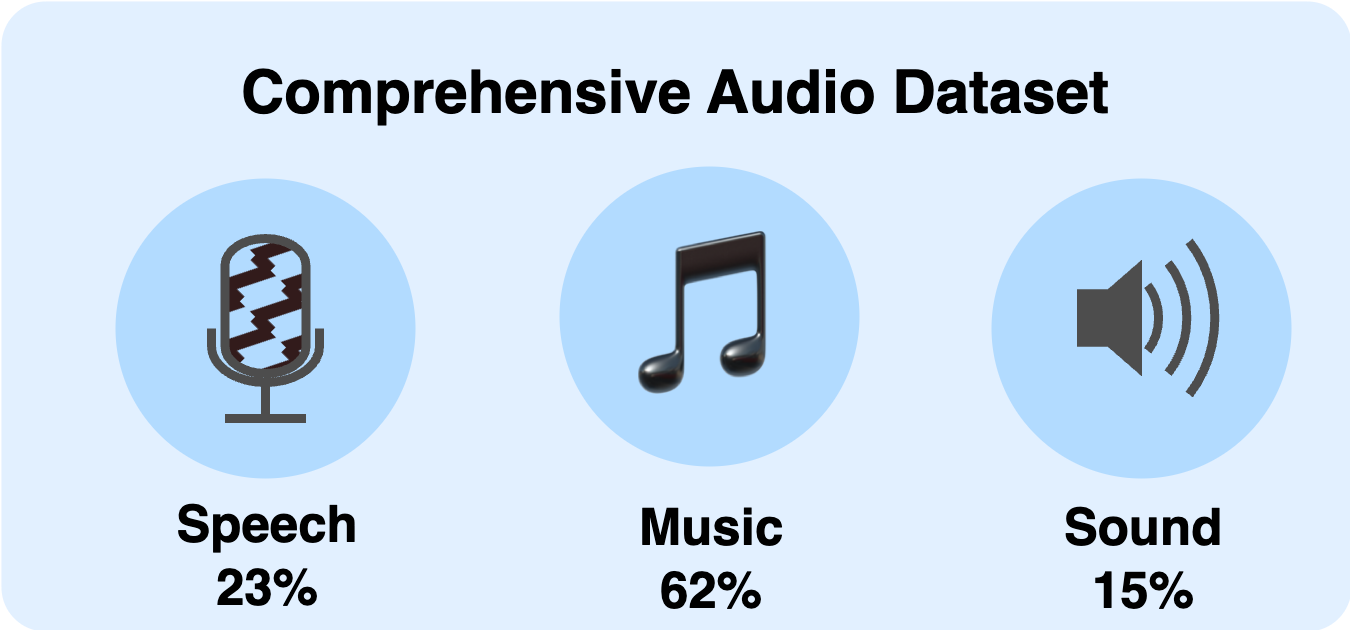}
    \caption{AudioCodecBench data distribution overview.}
    \label{fig: dataset}
\end{figure}

\begin{figure*}[t]
\centering
\includegraphics[width=1.0\textwidth]{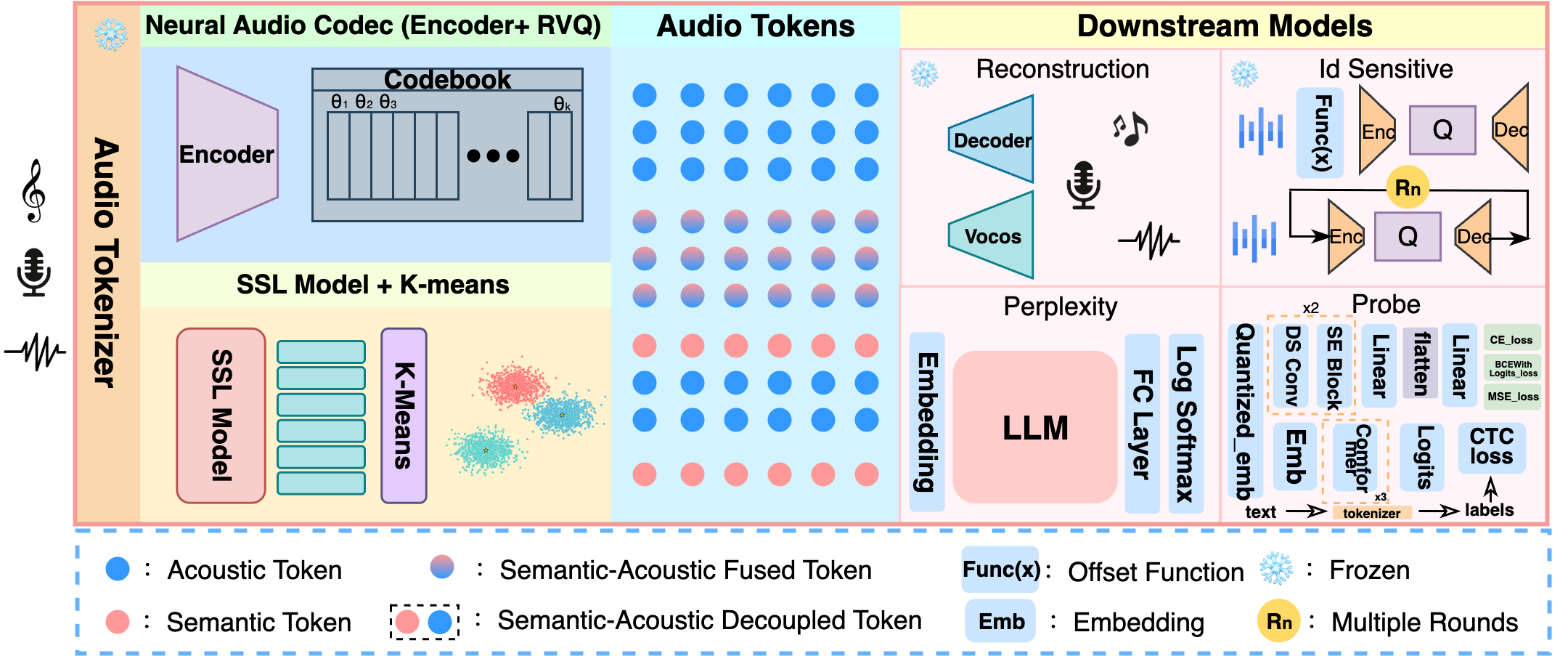} 
\caption{The proposed AudioCodecBench framework. Users provide pre-trained codec and obtain token-level outputs through encoding and quantization. Different types of tokens are input into different evaluation task components for multi-dimensional task evaluation.}
\label{overview}
\end{figure*}

The core task of Large Language Models (LLMs) is to predict next token in a sequence. This mechanism requires that its input must be a series of discrete tokens. Therefore, researchers always adopt the discrete quantization methods~\nocite{i15, i16}\citep{i15, i16}. These methods aim to approximate a large, continuous vector space with a finite, discrete set of representative vectors, mapping high-dimensional continuous signals in a finite codebook. Therefore, the signal can be translated effectively into token sequences that LLMs can understand and generate. These discrete methods function as a clustering process to generate codebook indices. Whether these indices represent semantics or acoustics depends on the encoder. However, despite growing research on discrete tokens, there is still no unified framework to evaluate and compare the performance of different token types.

To address these shortcomings, this paper introduces a systematic, multi-dimensional benchmark for codec evaluation. This benchmark comprehensively assesses codec capabilities across four key experiments: \textbf{Reconstruction}, to assess audio reconstruction fidelity; \textbf{ID Sensitivity}, to evaluate codebook ID stability under noisy conditions; \textbf{Perplexity}, to measure the impact of different token sequences on Large Models (LMs) modeling and \textbf{Probe}, to evaluate downstream task performance. The distribution of datasets is illustrated in Figure \ref{fig: dataset}. We hope that this benchmark will offer a more comprehensive comparison of various audio tokenization methods. Our contributions include the following:
\begin{itemize}
\item We provide suitable definitions of semantic and acoustic features. And base on their combination further define fused features and decoupled features.
\item We evaluate four features across a variety of tasks in our benchmark. This benchmark considers multiple evaluation metrics, and also covers three audio domains: speech, music and sound.
\item We explor the correlation between various task metrics and perplexity.
\end{itemize}

\begin{table*}
\centering
\setlength{\tabcolsep}{3.8pt} 
\begin{tabular}{c|ccccccc}
\toprule
\textbf{Feature Type} & \textbf{Model} & \textbf{\makecell{Sample \\  Rate}} & \textbf{\#Codebooks} & \textbf{\makecell{Codebook \\ Size}} & \textbf{\#Params} & \textbf{\makecell{Bitrate \\ (kbps)}} & \textbf{\makecell{Token \\ Rate}}\\
\midrule
\multirow{3}{*}{\textbf{Acoustic}} 
& DAC & 24kHz& 8 & 1024 & 74.7M & 6kbps & 75 \\
& EnCodec & 24kHz& 8 & 1024 & 14.9M & 6kbps & 75\\
& WavTokenizer & 24kHz& 1 & 4096 & 103M & 0.48kbps & 40\\
\midrule
\multirow{2}{*}{\textbf{Semantic}} 
& HuBERT & 16kHz& - & - & 94.4M & - & 50 \\
& Qwen2Audio & 16kHz& - & - & 636M & - & 25 \\
\midrule
\multirow{4}{*}{\textbf{Semantic and Acoustic Fused}} 
& SpeechTokenizer & 16kHz& 8 & 1024 & 80.9M & 5.33kbps & 50 \\
& Mimi & 24kHz& 8 & 2048 & 39.4M & 1.1kbps & 12.5 \\
& XCodec & 16kHz& 8 & 1024 & 123M & 4kbps & 50 \\
& YuE & 16kHz& 8 & 1024 & 123M & 4kbps & 50 \\
\midrule
\multirow{1}{*}{\textbf{Semantic and Acoustic Decoupled}} 
& SemantiCodec & 16kHz& 2 & 8192 & 507M & 1.3kbps & 100 \\
\bottomrule
\end{tabular}
\caption{The relevant feature types and parameters of the audio codecs and SSL models.}
\label{tab:feature}
\end{table*}

\section{Related Work}

\subsection{Audio Representation}
SpeechTokenizer~\nocite{i10}\citep{i10} distinguishes between ``Semantic token" and ``Acoustic token". Semantic token originates from SSL models like BEST-RQ~\nocite{r20}\citep{r20}, HuBERT~\nocite{i2}\citep{i2}, Wav2Vec2~\nocite{i3}\citep{i3} and WavLM~\nocite{r6}\citep{r6}. These models typically employ BERT-like structures and MLM loss to capture global contextual information, and it is often assume that semantics can be equated with performance on the Automatic Speech Recognition (ASR) task. However, we think that semantics is not only responded by ASR performance. In contrast, acoustic tokens are generated by codecs like EnCodec~\nocite{i7}\citep{i7}, SoundStream and DAC~\nocite{r5}\citep{r5} employ VQ-VAE driven by reconstruction loss to achieve high-fidelity reconstruction. This concept of audio representation provides a foundation for systematically analyzing the information types of discrete tokens.

To leverage the strengths of both token types, subsequent research explores different paradigms. SemantiCodec~\nocite{r12}\citep{r12} and XY-Tokenizer~\nocite{r21}\citep{r21} employs a dual-encoder architecture to decouple acoustic and semantic tokens by reconstruction loss and k-means clustering. In contrast, models like XCodec~\nocite{r10}\citep{r10}~\nocite{r11}\citep{r11} directly concatenate the two token types at the feature level. Meanwhile, SpeechTokenizer and Mimi~\nocite{r7}\citep{r7} introduce a “semantic distillation” approach. It uses an SSL model to guide the encoder of codec so that its discrete tokens carry both acoustic and semantic content in the first codebook. With the development of these different representation methods, establishing a fair and comprehensive evaluation becomes a significant challenge.

\subsection{SSL and Codec Benchmark}

Evaluation of discrete audio representations presents a diverse challenge. SSL benchmarks like SUPERB~\nocite{r1}\citep{r1} and MARBLE~\nocite{i8}\citep{i8} respectively evaluate representation performance on downstream tasks in the domains of speech and Music Information Retrieval. HEAR~\nocite{r2}\citep{r2} further extends the downstream tasks to multiple domains of speech, environment sounds and music. Similar to HEAR, ARCH~\nocite{r13}\citep{r13} introduces diverse datasets and offers a more extensible cross-domain evaluation framework than HEAR. However, a common limitation of these benchmarks is that they focus on downstream tasks, ignoring other evaluation aspects such as audio reconstruction and LM perplexity. Other methods of evaluation aspects like Code Drift~\nocite{i11}\citep{i11} evaluates the stability of multi-round reconstruction, while Codec-SUPERB~\nocite{r3}\citep{r3} evaluates reconstruction fidelity. DASB~\nocite{r4}\citep{r4} systematically probes discrete tokens in speech tasks.


To consolidate these diverse evaluation methods, researchers develop comprehensive toolkit like VERSA~\nocite{r14}\citep{r14}, and compile survey~\nocite{r17}\citep{r17} to integrate existing methods within a unified framework. However, these evaluation methods typically evaluate the performance of discrete tokens from diverse tasks. As a result, they do not define the different types of information of semantic and acoustic. And exploring the different types connect to different tasks. Therefore, there is an urgent need to bridge this gap.


This paper first establishes a suitable definition of ``semantic" that \textbf{must be strictly described by text}. Based on this, this paper further defines four different information types and compares the performance of discrete tokens of these four information types under different tasks. Through comprehensive experimental analysis, we explore the information types of various discrete tokens, providing insights to support the design of more effective audio representations.

\begin{table*}
\centering
\setlength{\tabcolsep}{1pt} 
\renewcommand{\arraystretch}{1.2}
\begin{tabular}{c|ccc}
\toprule
\textbf{Audio Type} & \textbf{Task} & \textbf{Dataset} & \textbf{Metric} \\
\midrule
\multirow{13}{*}{\textbf{Music}} & \multirow{1}{*}{Genre Classification(GC)} 
& GTZAN~\nocite{dataset4}\citep{dataset4} & Accuracy \\
\cline{3-4} 
& Key Detection(KD) & GiantSteps Key~\nocite{dataset1}\citep{dataset1} & Accuracy \\
\cline{3-4} 
& \multirow{2}{*}{Emotion Detection(ED)} & Emomusic~\nocite{dataset5}\citep{dataset5}  & $\mathrm{R}^2_{\mathrm{Valence}}$ \& $\mathrm{R}^2_{\mathrm{Arousal}}$ \\
&  & MTG MoodTheme~\nocite{dataset3}\citep{dataset3} & ROC-AUC \& PR-AUC/AP \\
\cline{3-4} 
& Vocal Technique Detection(VTD)& VocalSet~\nocite{dataset7}\citep{dataset7} & Accuracy \\
\cline{3-4} 
& Pitch Classification(PC) & NSynth~\nocite{dataset6}\citep{dataset6} & Accuracy \\
\cline{3-4} 
& \multirow{2}{*}{Music Tagging(MT)} & MagnaTagATun~\nocite{dataset2}\citep{dataset2} & ROC-AUC \& PR-AUC/AP \\
&  & MTG Top50~\nocite{dataset3}\citep{dataset3} & ROC-AUC \& PR-AUC/AP \\
\cline{3-4} 
& \multirow{2}{*}{Instrument Classification(IC)} & NSynth~\nocite{dataset6}\citep{dataset6} & Accuracy \\
&  & MTG Instrument~\nocite{dataset3}\citep{dataset3} & ROC-AUC \& PR-AUC/AP \\
\cline{3-4} 
& Singer Identification(SI) & VocalSet~\nocite{dataset7}\citep{dataset7} & Accuracy \\
\midrule
\multirow{2}{*}{\textbf{Speech}} 
& Automatic Speech Recognition(ASR) & Common Voice~\nocite{dataset14}\citep{dataset14}  & WER,CER \\
\cline{3-4} 
& Emotion Detection(ED) & MELD~\nocite{dataset10}\citep{dataset10} & Accuracy \\
\midrule
\multirow{2}{*}{\textbf{Sound}} & Vocal Sound Classification(VSC) & VocalSound~\nocite{dataset11}\citep{dataset11} & Accuracy \\
\cline{3-4} 
& Environmental Sound Classification(ESC) & ESC-50~\nocite{dataset12}\citep{dataset12} & Accuracy \\
\bottomrule
\end{tabular}
\caption{The task, dataset and evaluation metric for the downstream probe.The following text will use abbreviations to replace the full names of various tasks, datasets, and evaluation materials. Dataset-related GiantSteps Key: GS, Emomusic: EMO, MTG MoodTheme: MTGMT, VocalSet: VST, NSynth: NS, MagnaTagATun: MTT, MTG Top50: MTGT, MTG Instrument: MTGI, Common Voice: CV, VocalSound: VSD. Material-related ROC-AUC \& PR-AUC: RA.}
\label{tab:audio_tasks}
\end{table*}

\section{Evaluation Framework}

\subsection{Overall Architecture}

In the reconstruction task, we process an original audio signal through the encoder, quantizer, and decoder pipeline to reconstruct waveform, and use metrics like Perceptual Evaluation of Speech Quality (PESQ)~\nocite{e7}\citep{e7}, Short-Time Objective Intelligibility (STOI)~\nocite{e8}\citep{e8} to evaluate the codec’s ability to encode acoustic details; while using Word Error Rate (WER) and Character Error Rate(CER) to evaluate semantic preservation in acoustic details. The codec with higher metric scores is considered to have tokenization more focused on accurately reconstructing acoustic details.

The ID sensitivity experiment consists of two subtasks, as shown in the upper right section of the downstream model in Figure \ref{overview}. The first task is multi-round reconstruction, we use the output of the $(n)th$ round as the input for the $(n+1)th$ round. The second task is the temporal shift stability experiment. We simulate signal phase shift by introducing millisecond-level time shifts into the original audio, and reconstruct this shifted audio. We define \textbf{ID sensitivity} as the stability of discrete tokens under noise interference. For both subtasks, we calculate the unchanged rate of IDs in the same codebook after the process to evaluate the representation's robustness. Higher stability indicates lower ID sensitivity, and conversely, lower stability indicates higher ID sensitivity.

For the perplexity experiment, we extract the sequence of discrete IDs from the codec, then train a small LM using the Cross-Entropy loss to predict next audio-only tokens. As shown in the lower left section of the downstream model in Figure \ref{overview}. We use the perplexity of this LM as the evaluation metric to evaluate the adaptability of the discrete ID sequence for LM modeling. A lower perplexity indicates that the sequence is more amenable to LM modeling and also implies that it may contain richer semantics.

In the downstream probe model, we design two structures to evaluate the generalization of discrete tokens through various downstream tasks. As shown in the lower right section of the downstream model in Figure \ref{overview}. The first is a lightweight network composed of SE‑Blocks~\nocite{e9}\citep{e9} (channel attention) and depthwise separable convolutions~\nocite{e10}\citep{e10}. This network compresses both the temporal and feature dimensions of the embedding after quantization and then makes predictions using task-specific heads. For the ASR task, we design a different approach to measure the alignment between the representation and text. The extracted discrete IDs are fed through an embedding layer into a three-layer Conformer network~\nocite{e5}\citep{e5}, and the model is trained end-to-end using the Connectionist Temporal Classification (CTC) loss~\nocite{e6}\citep{e6}.

\subsection{Audio Feature Classification}
We review existing definitions of audio representations (acoustic and semantic), but find these definitions fail to cover the current diverse features. \textbf{Therefore, we propose that a semantic feature must be strictly describable by text.} On this basis, we divide audio features into four categories.

\textbf{1) Acoustic feature}: The discrete feature \textbf{cannot be described by text}. These features originate from codecs optimized for reconstruction, representing the quantized encoding of acoustic details, such as environmental noise, vocal fold vibration and air vibration.

\textbf{2) Semantic feature}: The discrete feature extracted from MLM within SSL frameworks \textbf{must be strictly defined by text}. They aim to capture high-level and abstract information, such as the transcribed text of speech, the emotion or key of music and the human voice in music.

\textbf{3) Semantic-Acoustic fused feature}: The discrete features is \textbf{fused with both text-describable semantics and text-indescribable acoustic information}.  For instance, features representing a specific speaker's voice simultaneously contain textual content and unique acoustic details.

\textbf{4) Semantic-Acoustic decoupled feature}: The discrete features that \textbf{separates text-describable semantics and text-indescribable acoustic information into independent codebooks}. For the same audio clip of `Hello', it outputs two independent token streams: one representing the text-describable information `Hello', and the other representing text-indescribable acoustic details such as the speaker’s unique acoustic signature.

Based on the definitions of the four feature classes, the codecs and SSL models evaluated in this paper are classified accordingly. Table \ref{tab:feature} provides a summary of these models, detailing the model feature types they generate and key technical specifications such as sample rate, bit rate, and token rate.

\begin{table*}
\centering
\begin{tabular}{lccccc}
\toprule
\textbf{Codec} & \textbf{PESQ$\uparrow$} & \textbf{Spk-Sim$\uparrow$} & \textbf{WER (GT/REC)$\downarrow$} & \textbf{CER (GT/REC)$\downarrow$} & \textbf{STOI$\uparrow$} \\
\midrule
DAC & \textbf{3.69} / \textbf{2.66} & \textbf{0.965} / - & 0.155 / 0.202 \textbar\  - / - & 0.09 / 0.125 \textbar\ - / - & \textbf{0.94} / \textbf{0.86} \\
EnCodec & 3.21 / 2.27 & 0.919 / - & 0.155 / 0.198 \textbar\  - / - & 0.09 / 0.114 \textbar\ - / - & 0.93 / 0.85 \\
Mimi & 2.77 / - & 0.928 / - & 0.155 / 0.287 \textbar\ - / - & 0.09 / 0.173 \textbar\   - / - & 0.88 / - \\
SemantiCodec & 2.64 / 1.32 & 0.907 / - & 0.155 / 0.318 \textbar\  - / - & 0.09 / 0.195 \textbar\ - / - & 0.86 / 0.60 \\
WavTokenizer & 2.17 / 1.14 & 0.743 / - & 0.155 / 0.494 \textbar\  - / - & 0.09 / 0.325 \textbar\ - / - & 0.83 / 0.49 \\
SpeechTokenizer & 2.97 / - & 0.924 / - & 0.155 / 0.216 \textbar\  - / - & 0.09 / 0.120 \textbar\ - / - & 0.89 / - \\
XCodec & 3.23 / 1.85 & 0.942 / - & \textbf{0.155 / 0.185} \textbar\  - / - & \textbf{0.09 / 0.106} \textbar\  - / - & 0.91 / 0.76 \\ 
YuE & 3.17 / 1.84 &  0.938 / - &  0.155 / 0.195 \textbar\  - / - &  0.09 / 0.113 \textbar\  - / -&  0.90 / 0.75 \\ 
\bottomrule
\end{tabular}
\caption{Reconstruction results of difference codecs in LibriTTS test-other dataset and GTZAN test dataset.}
\label{tab: codec_rec_eval_speech}
\end{table*}

\begin{figure*}[h]
    \centering
    \includegraphics[width=1.0\linewidth]{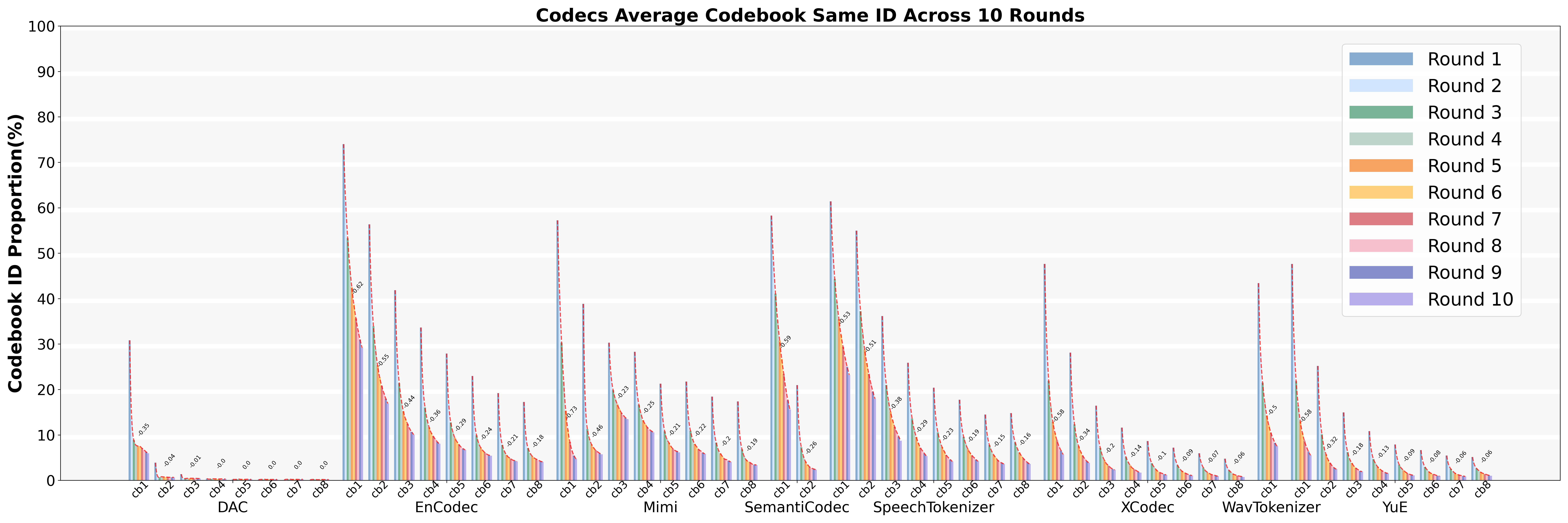}
    \caption{The percentage of the same ID in each codebook of the codecs after multi-round reconstruction, cb stands for codebook.}
    \label{fig: codec_multiround_id}
\end{figure*}

\begin{figure*}[h]
    \centering
    \includegraphics[width=1.0\linewidth]{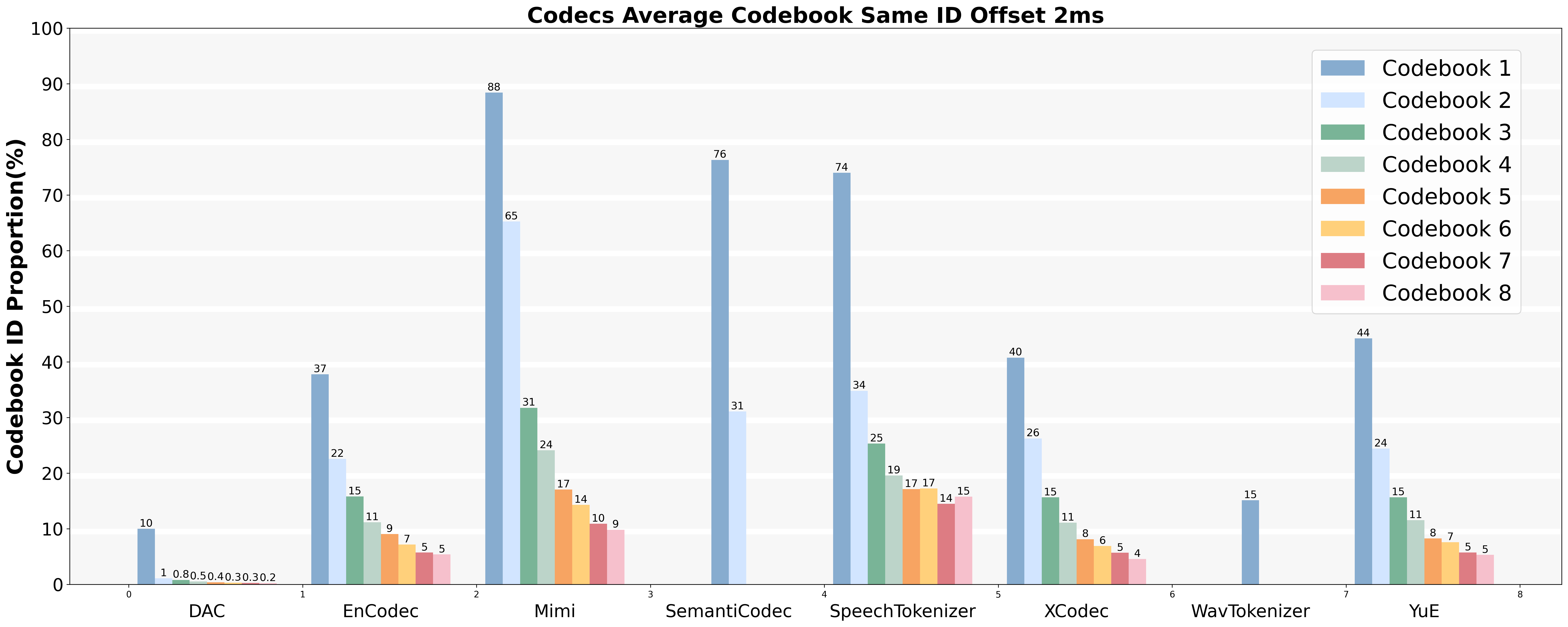}
    \caption{The proportion of identical IDs in each codebook of the codecs after time shift processing and reconstruction.}
    \label{fig: codec_os2_same_id}
\end{figure*}

\section{Experiments and Analysis}
We evaluate the performance of eight codecs and two SSL models. Their relevant attributes are listed in Table \ref{tab:feature}. We use the first 8 codebooks to
evaluate the performance of the multi-codebook codecs.

\subsection{Reconstruction}
We conduct reconstruction experiment on the LibriTTS test-other~\nocite{dataset9}\citep{dataset9} and GTZAN test datasets. In Table \ref{tab: codec_rec_eval_speech}, the results are rounded to the required precision for each metric. The left side of each metric result is the speech dataset result, and the right side is the music dataset result. Since Mimi and SpeechTokenizer are not trained on music datasets, they are not evaluated on music dataset experiments.

On the speech dataset, acoustic codecs such as DAC and EnCodec achieve the highest reconstruction fidelity. Codecs that integrate semantics like XCodec and YuE demonstrate the suboptimal performance, while WavTokenizer performs the worst. The result suggests that semantics may affect the reconstruction of acoustic details. Although WavTokenizer's discrete tokens are acoustic, its reconstruction quality is weak. We speculate that to balance compression bitrate and reconstruction quality, \textbf{small codebook size and few codebooks limit the variety of combinations for the discrete tokens, which weakens the ability of these tokens to capture acoustic details}. 

All reconstruction metrics are lower on the music dataset compared to the speech dataset. This is because music contains more intricate harmonic structures and richer dynamic variations than speech. Therefore, music is more difficult to model and reconstruct with high-fidelity. Notably, the performance of WavTokenizer and SemantiCodec decreases significantly. This result further highlights the limitations of small codebook size and the single or dual-codebook quantization strategies. \textbf{Small codebook size and few codebooks limit the possibility of token combinations to represent the acoustic details of music, thus reducing reconstruction fidelity}. In particular, WavTokenizer exhibits poor modeling capabilities for music, resulting in a decrease in subjective listening quality after reconstruction.

\begin{table*}
\centering
\setlength{\tabcolsep}{4.7pt} 
\begin{tabular}{lccccccccc}
\toprule
\textbf{Codec}  & \textbf{ppl$\downarrow$}  & \textbf{cb1\_ppl} & \textbf{cb2\_ppl} & \textbf{cb3\_ppl} & \textbf{cb4\_ppl} & \textbf{cb5\_ppl} & \textbf{cb6\_ppl} & \textbf{cb7\_ppl} & \textbf{cb8\_ppl}\\
\midrule
DAC & 247 / 194 & 21 / 29 & 147 / 123 & 218 / 152 & 315 / 213 & 396 / 2701 & 483 / 353 & 570 / 413 & 628 / 474 \\
EnCodec & 76 / 141 & 15 / 18 & 33 / 63 & 59 / 111 & 89 / 170 & 111 / 226 & 138 / 287 & 159 / 337 & 173 / 376 \\
WavTokenizer & 105 / 38 & 105 / 38 & - / - & - / - & - / - & - / - & - / - & - / - & - / - \\
XCodec & 30 / 48 & 10 / 20 & 13 / 20 & 20 / 32 & 31 / 51 & 42 / 65 & 51 / 75 & 62 / 87 & 71 / 100 \\
YuE & 29 / 46 & 9 / 18 & 16 / 29 & 20 / 30 & 29 / 48 & 39 / 60 & 51 / 75 & 55 / 83 & 54 / 76  \\
SpeechTokenizer & 14 / - & 2 / - & 6 / - & 12 / - & 18 / - & 22 / - & 25 / - & 29 / - & 31 / - \\
Mimi & 127 / - & 9 / - & 58 / - & 148 / - & 185 / - & 229 / - & 257 / - & 279 / - & 299 / - \\
SemantiCodec & \textbf{8} / \textbf{16} & 1 / 1 & 82 / 272 & - / - & - / - & - / - & - / - & - / - & - / - \\
\bottomrule
\end{tabular}
\caption{PPL results of different codecs in Emilia-EN dataset and MTG-Jamendo dataset, cb stands for codebook.}
\label{tab:ppl_speech}
\end{table*}

\subsection{ID sensitivity}
We evaluate ID sensitivity through multi-round $(n=10)$ reconstruction and time shift task. The results are shown in Figure \ref{fig: codec_multiround_id}. Detailed results for different codecs are shown in ~\autoref{Appendix_A}A. After multi-round reconstruction, the codebook IDs of all codecs shift compared to the first round. Codecs focusing on acoustic reconstruction show higher ID stability (\textbf{lower slope} of the decrease rate of the same ID). The result indicates that they can accurately reconstruct the signal, including some possible noise. In contrast, codecs that integrate semantics exhibit lower ID stability (\textbf{higher slope}). The result shows that these codecs are less sensitive to fitting noise during reconstruction and focus more on ensuring semantic. Although EnCodec generates tokens that are mainly acoustic, its multi-round reconstruction performance is similar to the codecs integrating semantics. This may be attributed to EnCodec’s integration of LSTM modules during encoding, which capture long-context dependencies, enhancing the stability of multi-round reconstruction.

Inspired by Code Drift~\nocite{i11}\citep{i11}, we select 2ms as the experimental setting for time shift task, the results are shown in Figure \ref{fig: codec_os2_same_id}. Detailed results for different codecs are shown in \autoref{Appendix_B}B. The result demonstrates that the token sequences of acoustic codecs are sensitive to temporal changes, as they focus on reconstruction and attempt to encode all acoustic details, including slight timing shifts. And codecs that integrate semantics focus more on stable content features, thus demonstrating greater robustness to slight timing shifts. \textbf{Codecs that integrate semantics outperform the acoustic codecs on the same ID ratio metric, which indicates that semantic-dominant tokens are more robust to slight timing shifts.}

\begin{table*}
\centering
\setlength{\tabcolsep}{3pt} 

\begin{tabular}{l|c|cccc|cccc|ccc|c|c|c|c} 
\toprule
\textbf{Task} & \multicolumn{1}{c|}{\textbf{GC}}  & \multicolumn{4}{c|}{\textbf{ED}} & \multicolumn{4}{c|}{\textbf{MT}} & \multicolumn{3}{c|}{\textbf{IC}} & \textbf{KD} & \textbf{VTD} & \textbf{PC} & \textbf{SI}\\
\midrule
\textbf{Dataset} & \textbf{GTZAN}  & \multicolumn{2}{c}{\textbf{EMO}} & \multicolumn{2}{c|}{\textbf{MTGMT}}  & \multicolumn{2}{c}{\textbf{MTT}} & \multicolumn{2}{c|}{\textbf{MTGT}} & \textbf{NS} & \multicolumn{2}{c|}{\textbf{MTGI}} & \textbf{GS} & \textbf{VST} & \textbf{NS} & \textbf{VST}\\
\textbf{Metrics} & \textbf{Acc$\uparrow$} & \textbf{$\mathrm{R}^2_{\mathrm{A}}$$\uparrow$} & \textbf{$\mathrm{R}^2_{\mathrm{V}}$$\uparrow$} & \textbf{AP$\uparrow$} & \textbf{RA$\uparrow$} & \textbf{AP$\uparrow$} & \textbf{RA$\uparrow$} & \textbf{AP$\uparrow$} & \textbf{RA$\uparrow$} & \textbf{Acc$\uparrow$} & \textbf{AP$\uparrow$} &\textbf{RA$\uparrow$} & \textbf{Acc$\uparrow$} & \textbf{Acc$\uparrow$} & \textbf{Acc$\uparrow$} & \textbf{Acc$\uparrow$} \\
\midrule
DAC & 0.58  & 0.47 & 0.06 & 0.08 & 0.65 & 0.20 & 0.79 & 0.14 & 0.69 & 0.60 & 0.11 & 0.64 & 0.09 & 0.38 & 0.47 & 0.42\\
EnCodec & 0.57 & 0.47 & 0.07 & 0.06 & 0.64 & 0.18 & 0.76 & 0.14 & 0.70 & 0.54 & 0.10 & 0.62 & 0.10 & 0.30 & 0.55 & 0.30 \\
WavTokenizer & 0.42  & 0.46 & 0.07 & 0.06 & 0.63 & 0.17 & 0.74 & 0.14 & 0.70 & 0.54 & 0.11 & 0.64 & 0.09 & 0.29 & 0.44 & 0.13\\
SemantiCodec & \textbf{0.70}  & 0.51 & \textbf{0.32} & 0.10 & \textbf{0.72} & 0.32 & \textbf{0.88} & \textbf{0.23} & \textbf{0.80} & \textbf{0.66} & 0.15 &\textbf{0.72} & 0.34 & 0.45 & 0.76 & 0.34\\
XCodec & 0.66 & 0.55 & 0.14 & 0.10 & 0.71 & \textbf{0.32} & 0.87 & 0.22 & 0.78 & 0.64 & \textbf{0.16} & 0.71 & \textbf{0.46} & 0.57 & \textbf{0.91} & \textbf{0.54}\\
YuE & 0.67 & \textbf{0.57} & 0.16 & \textbf{0.10} & 0.71 & 0.32 & 0.87 & 0.19 & 0.76 & 0.62 & 0.13 & 0.70 & 0.45 & \textbf{0.59} & 0.90 & 0.52\\
\bottomrule
\end{tabular}
\caption{The results of various detection tasks performed by the codecs across different music datasets.} 
\label{tab: probe_music_eval} 
\end{table*}

\begin{table}
\centering
\setlength{\tabcolsep}{2.5pt} 
\begin{tabular}{l|cc|c|c|c} 
\toprule
\textbf{Task} & \multicolumn{2}{c|}{\textbf{ASR}}& \textbf{\makecell{VSC}} & \textbf{\makecell{ESC}} & \textbf{ED} \\
\midrule
\textbf{Dataset} & \multicolumn{2}{c|}{\textbf{CV}} & \textbf{VSD} & \textbf{ESC-50} & \textbf{MELD} \\
\textbf{Metrics} & \textbf{WER$\downarrow$} &\textbf{CER$\downarrow$} & \textbf{Acc$\uparrow$} & \textbf{Acc$\uparrow$} & \textbf{Acc$\uparrow$} \\
\midrule
DAC & 0.53 & 0.23 & 0.54 & 0.33 & 0.48 \\
EnCodec & 0.50 & 0.21 & 0.57 & 0.28 & 0.48 \\
WavTokenizer & 0.58 & 0.29 & 0.52 & 0.14 & 0.48 \\
SemantiCodec & 0.49 & 0.20 & 0.72 & 0.62 & 0.48 \\
Mimi & \textbf{0.44} & \textbf{0.17} & 0.83 & 0.34 & 0.48 \\
SpeechTokenizer & 0.47 & 0.19 & 0.78 & 0.67 & 0.50 \\
XCodec & 0.47 & 0.19 & 0.73 & 0.64 & 0.49  \\
YuE & 0.47 & 0.19 & 0.78 & 0.64 & 0.52 \\
HuBERT & - & - & 0.88 & 0.53 & 0.50 \\
Qwen2Audio & - & - & \textbf{0.95} & \textbf{0.98} & \textbf{0.59} \\
\bottomrule
\end{tabular}
\caption{The results of various detection tasks performed by the codecs and SSL models across different speech datasets.} 
\label{tab: probe_speech_eval} 
\end{table}

\subsection{Perplexity}
We train a 100M LM using Qwen2 architecture~\nocite{r9}\citep{r9} from scratch to evaluate the modeling efficiency of codecs via validation set perplexity (PPL). For multi‑codebook codecs, we apply a parallel evaluation~\nocite{e1}\citep{e1} to compute PPL for each codebook. To ensure a fair comparison, the PPL values are normalized, and the final PPL is calculated using a mean loss. Because PPL scores are directly influenced by the codebook size; larger codebooks typically result in higher PPL. Therefore, we normalize all values to a reference codebook size of 1024. The calculation is as follows:

\begin{equation}
\text{PPL} = \frac{\exp(\mathcal{L}_{CE})}{S_{\text{cb}} / 1024}
\end{equation}

\noindent
where $\mathcal{L}_{CE}$ is the average cross-entropy loss calculated over the entire token sequence, and $S_{\text{cb}}$ denotes the codec codebook size. The training runs for 100k steps on 8 NVIDIA A6000 GPUs using the Emilia‑EN~\nocite{dataset13}\citep{dataset13} and MTG‑Jamendo datasets. Table \ref{tab:ppl_speech} presents the results, rounded to the nearest integer. The left side of PPL metric result is the speech dataset result, and the right side is the music dataset result. Since Mimi and SpeechTokenizer are not trained on music datasets, they are not evaluated on music dataset experiments.

On the speech dataset, codecs that integrate semantics achieve better results than acoustic codecs. This result demonstrates that \textbf{semantic tokens are easier for LMs to model}. Analysis of the multi-codebook codecs' results shows that earlier codebooks have lower PPL, which provides strong support for the conclusion that semantics is beneficial for LM modeling. Although EnCodec mainly generates acoustic tokens, it achieves unexpectedly low PPL. Mimi uses a semantic teacher to guide its first quantizer, but it fails to achieve the performance of other codecs that integrate semantics. The exact reasons behind these unusual results are still unknown and need further exploration.

The PPL is higher on the music dataset compared to the speech dataset, the finding that is consistent with human intuition. This is because music involves multiple instruments and complex temporal structures. These factors create a larger variety of possible token combinations, making their distribution much sparser than in speech. However, the PPL values for DAC and WavTokenizer on the music dataset are unexpectedly lower than on the speech dataset. We speculate that this is because DAC and WavTokenizer were  trained on the MTG-Jamendo dataset but not on the Emilia-EN dataset, so their ppl results are different from other codecs.


\subsection{Probe}
In the downstream probe tasks, to ensure fair results, all experiments are conducted under the same computational budget. For the ASR task, we select Speech2Text~\nocite{e3,e4}\citep{e3,e4} as the text tokenizer. The related tasks, datasets, and evaluation metrics are shown in Table \ref{tab:audio_tasks}. Detailed introductions are shown in \autoref{Appendix_D}D.

The results of the music probe task are shown in Table \ref{tab: probe_music_eval}. The visualized result is shown in Figure \ref{music_probe} in \autoref{Appendix_C}C. In the ED task, SemantiCodec's performance on Valence prediction is the best. Arousal is more strongly associated with acoustic features, while Valence is more strongly associated with semantic content~\nocite{c1}\citep{c1}. This is consistent with our results. Tasks such as MT, GC and KD involve high-level musical structures, SemantiCodec shows advantages in these tasks. Meanwhile, XCodec and SemantiCodec also achieved better performance in IC and PC tasks, which closely related to symbolic music information. These tasks share the common feature that their labels (e.g., “Pop,” “A major”) can be strictly described by text or symbols, with a correspondence between musical content and labels. We refer to these tasks as semantic-driven tasks. Therefore, in these tasks, semantic codecs show better performance than acoustic codecs. These results also validate our definition of “semantic,” proving that \textbf{introducing semantics can effectively capture high-level, symbolizable information in music}.

The results of the speech and sound probe tasks are shown in Table \ref{tab: probe_speech_eval}. The visualized result is shown in Figure \ref{fig17} in \autoref{Appendix_C}C. The SSL models achieve the best performance. Codecs that integrate semantics demonstrate the suboptimal performance. Acoustic codecs perform the worst. WavTokenizer achieves the lowest performance. In the ASR task, codecs that explicitly introduce semantics generally achieve better WER/CER scores than acoustic codecs. In the VSC task, codecs that integrate semantics show outstanding performance. It further suggests that timbre information may be effectively retained and utilized in representations that contain both semantics and acoustics. In the ED task, the performance of different codecs is relatively balanced. This suggests that the emotion-related features required for this specific task can be fully fitted by codecs. 

\begin{table}[htbp]
\centering
\setlength{\tabcolsep}{3.7pt} 
\renewcommand{\arraystretch}{0.99}
\begin{tabular}{cccc}
\toprule
\textbf{Task} & \textbf{Dataset Type} & \textbf{Metric} & $\textbf{r}$ \\
\midrule
\multirow{5}{*}{\textbf{Reconstruction}} 
& \multirow{5}{*}{\textbf{Speech}} & \textbf{WER$_{REC}$} & 0.06\\
&  & \textbf{CER$_{REC}$} & 0.1 \\
&  & \textbf{PESQ} & -0.35 \\
&  & \textbf{Spk\_Smi} & -0.05 \\
&  & \textbf{STOI} & -0.35 \\
\midrule
\multirow{2}{*}{\textbf{ID sensitivity}} 
& \multirow{2}{*}{\textbf{Speech}} & \textbf{MRC} & 0.52 \\
&  & \textbf{OS} & 0.44 \\
\midrule
\multirow{20}{*}{\textbf{Probe}} 
& \multirow{5}{*}{\textbf{Speech}} & \textbf{WER$_{CTC}$} & 0.37 \\
& & \textbf{CER$_{CTC}$} & 0.36 \\
& & \textbf{VSC$_{ACC}$} & 0.55 \\
& & \textbf{ESC}$_{ACC}$ & 0.67 \\
& & \textbf{ED}$_{ACC}$ & 0.47 \\
\cmidrule(l){2-4}
& \multirow{16}{*}{\textbf{Music}} & \textbf{GC}$_{ACC}$ & 0.2 \\
&   & \textbf{ED}$_{\mathrm{R}^2_{\mathrm{A}}(EMO)}$ & 0.5 \\
&   & \textbf{ED}$_{\mathrm{R}^2_{\mathrm{V}}(EMO)}$ & 0.65 \\
&   & \textbf{ED}$_{AP(MTGMT)}$ & 0.43 \\
&   & \textbf{ED}$_{RA(MTGMT)}$ & 0.58 \\
&   & \textbf{MT}$_{AP(MTT)}$ & 0.59 \\
&   & \textbf{MT}$_{RA(MTT)}$ & 0.49 \\
&   & \textbf{MT}$_{AP(MTGT)}$ & 0.68 \\
&   & \textbf{MT}$_{RA(MTGT)}$ & 0.73 \\
&   & \textbf{IC}$_{ACC(NS)}$ & 0.39 \\
&   & \textbf{IC}$_{AP(MTGI)}$ & 0.65 \\
&   & \textbf{IC}$_{RA(MTGI)}$ & 0.71 \\
&   & \textbf{KD}$_{ACC}$ & 0.62 \\
&   & \textbf{VTD}$_{ACC}$ & 0.41 \\
&   & \textbf{PC}$_{ACC}$ & 0.56 \\
\bottomrule
\end{tabular}
\caption{Pearson correlation coefficient between PPL and metrics from various evaluation tasks.}
\label{tab:correlation}
\end{table}

In order to explore the impact of various metrics on LM modeling, we calculate the Pearson correlation coefficients between various task metrics and PPL. We aim to reveal which metrics or audio features are more beneficial for LM modeling. The correlation coefficient is performed on the results are shown in Table \ref{tab:correlation}. PPL is positively correlated with CTC probe task metrics and very weak correlated with reconstructed WER/CER metrics, which demonstrates that tokens rich in semantic content are easier for LMs to model. However, it shows a negative correlation with objective acoustic reconstruction metrics, indicating that overfitting acoustic details may increase the difficulty of LM modeling. The ID sensitivity metrics show a positive correlation with PPL, which indicates that introducing semantics can bring more stable ID patterns, thereby benefiting the modeling of LMs.

\section{Conlusion}
This paper presents a comprehensive, fair and highly reusable evaluation framework for codecs. We first redefine “acoustic” and “semantic” features: \textbf{semantic features must be strictly described by text}. Based on this classification, our benchmark systematically evaluates the performance of different discrete tokens across multiple tasks, and breaking the limitation of measuring semantics through ASR performance. Experimental results not only show the potential applications of various representations in MLLMs but also point to a new research direction: training better audio-semantic models by aligning text modality. We are committed to providing an open and fair benchmark and hope to attract more researchers to participate, jointly advancing the field of audio representation learning.

\bibliography{aaai2026}

\clearpage

\appendix

\section{Appendix A: Multi-round Reconstruction results of different codecs}
\label{Appendix_A}

\begin{figure}[H]
\centering
\includegraphics[width=1.0\columnwidth]{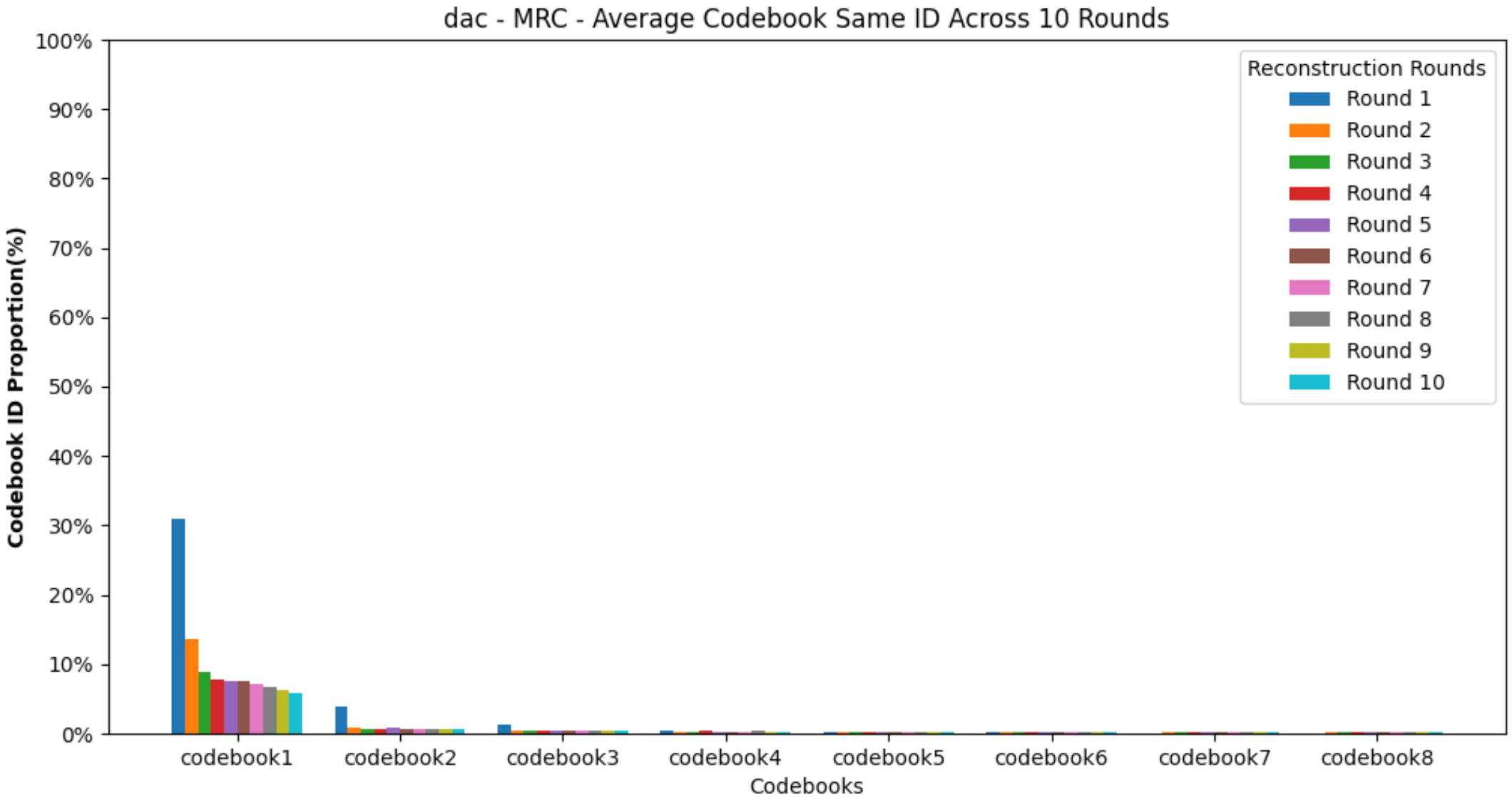} 
\caption{Multi-round Reconstruction results of DAC.}
\label{fig1}
\end{figure}

\vspace{-16pt}

\begin{figure}[H]
\centering
\includegraphics[width=1.0\columnwidth]{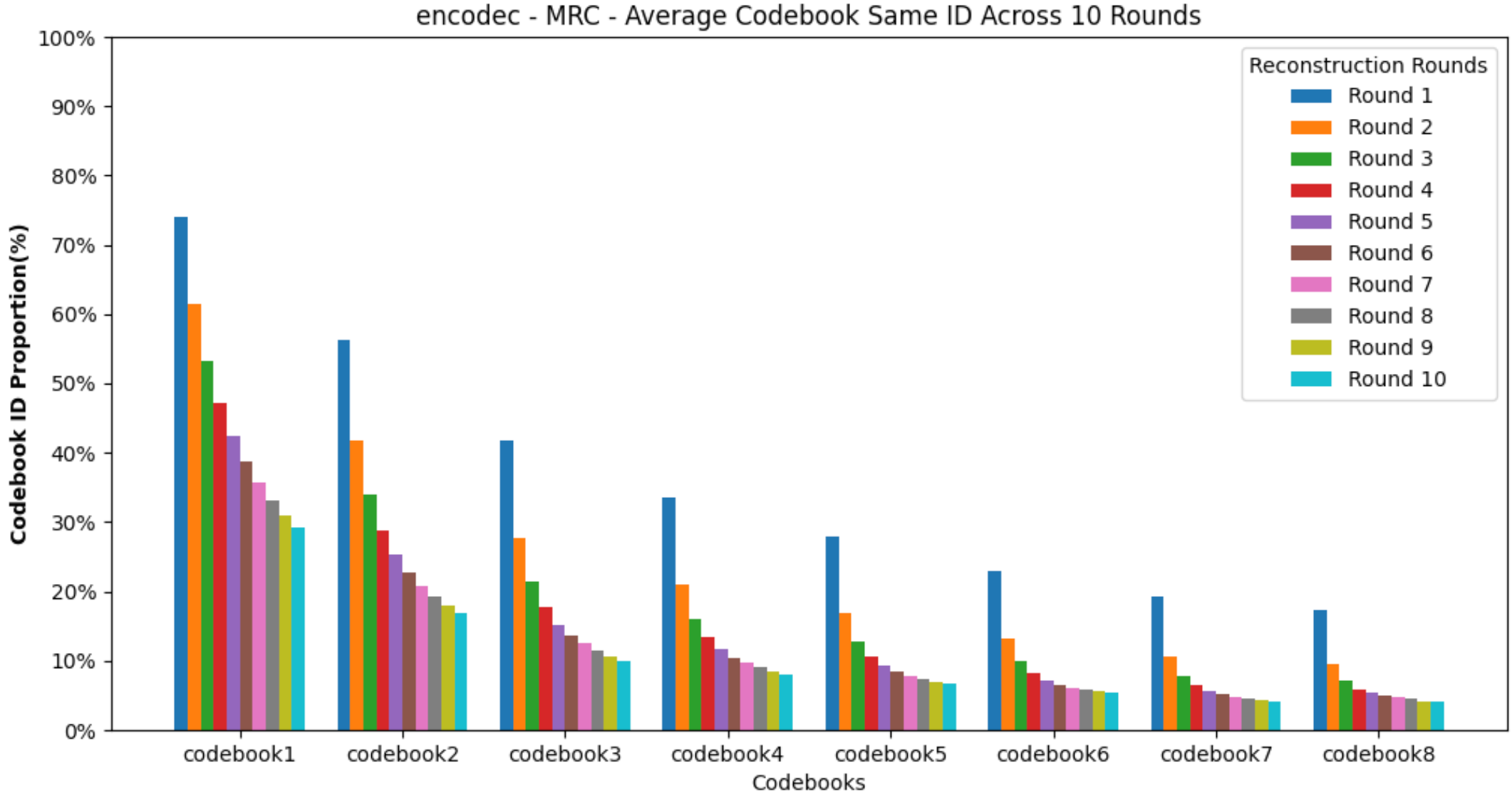} 
\caption{Multi-round Reconstruction results of EnCodec.}
\label{fig2}
\end{figure}

\vspace{-16pt}

\begin{figure}[H]
\centering
\includegraphics[width=1.0\columnwidth]{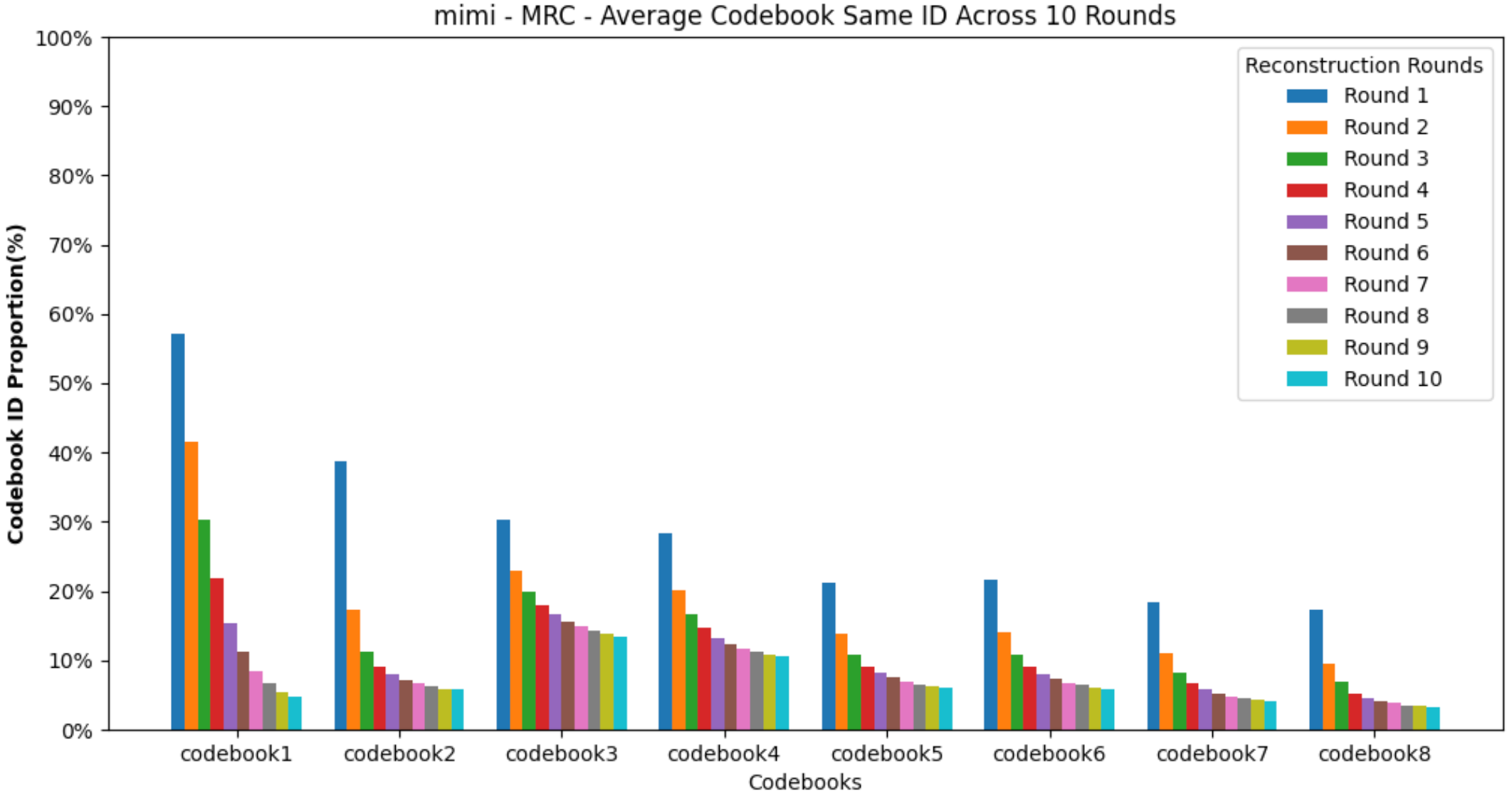} 
\caption{Multi-round Reconstruction results of Mimi.}
\label{fig3}
\end{figure}

\vspace{-16pt}

\begin{figure}[H]
\centering
\includegraphics[width=1.0\columnwidth]{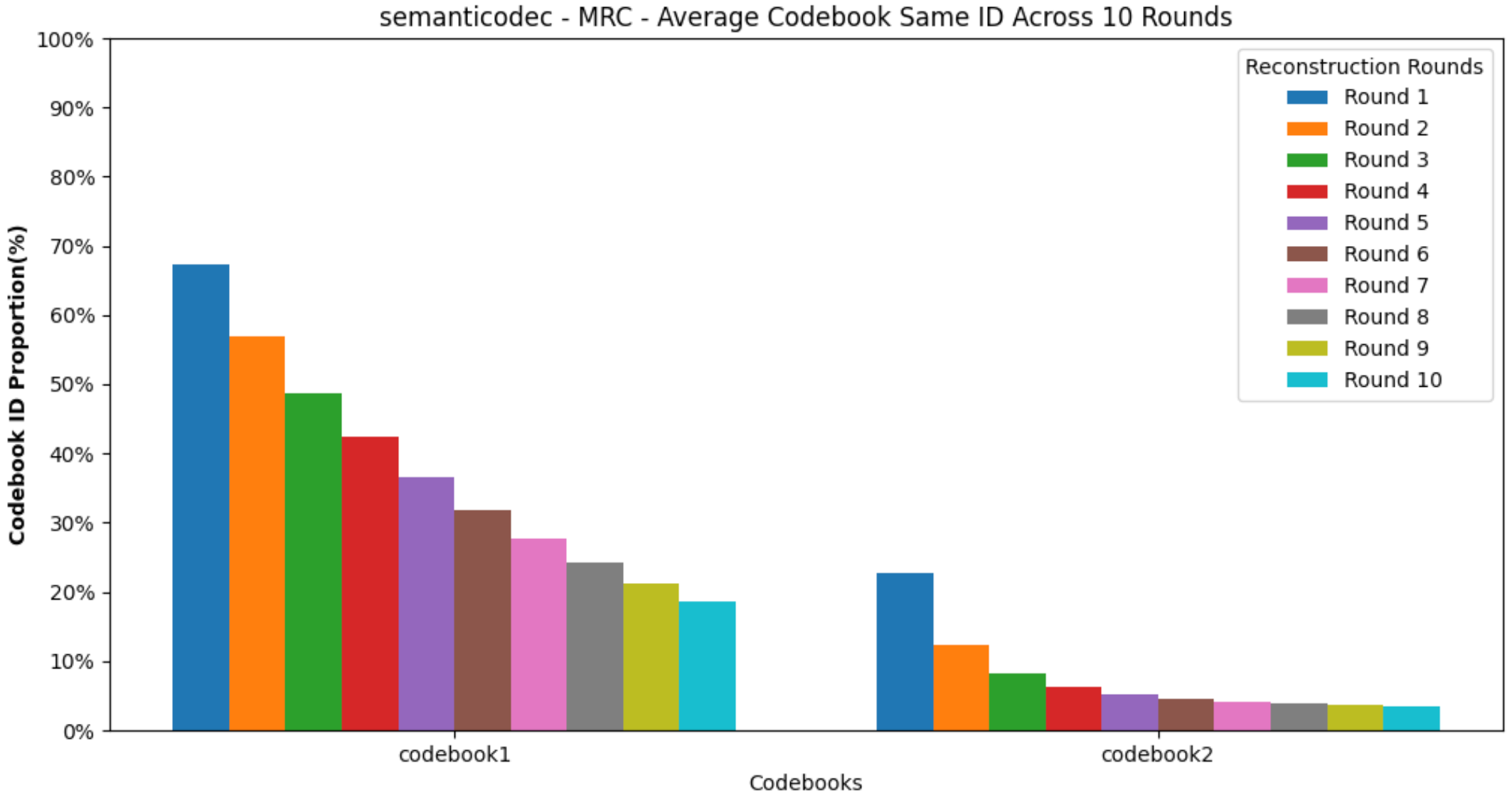} 
\caption{Multi-round Reconstruction results of SemantiCodec.}
\label{fig4}
\end{figure}

\vspace{-16pt}

\begin{figure}[H]
\centering
\includegraphics[width=1.0\columnwidth]{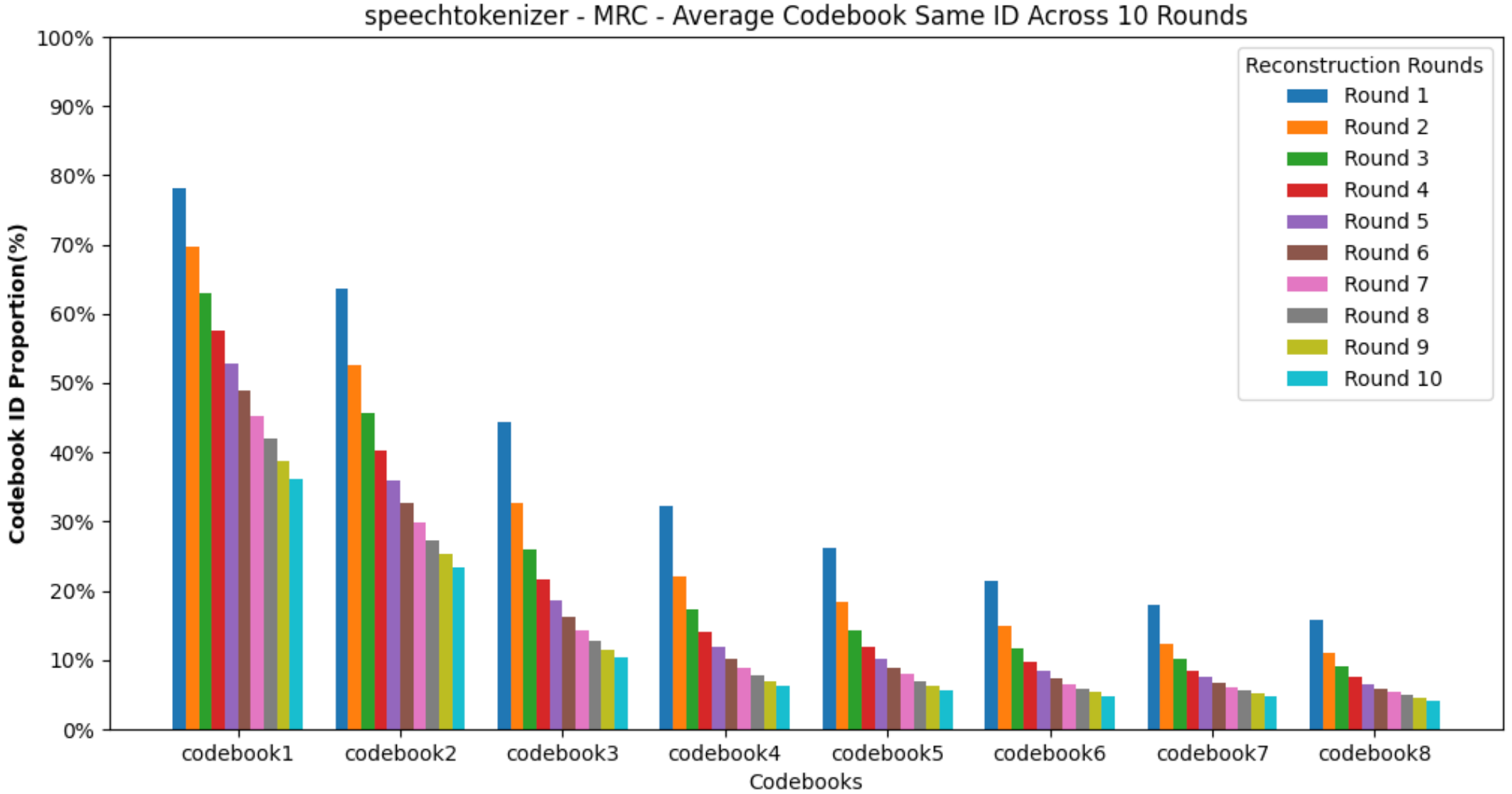} 
\caption{Multi-round Reconstruction results of SpeechTokenizer.}
\label{fig5}
\end{figure}

\vspace{-16pt}

\begin{figure}[H]
\centering
\includegraphics[width=1.0\columnwidth]{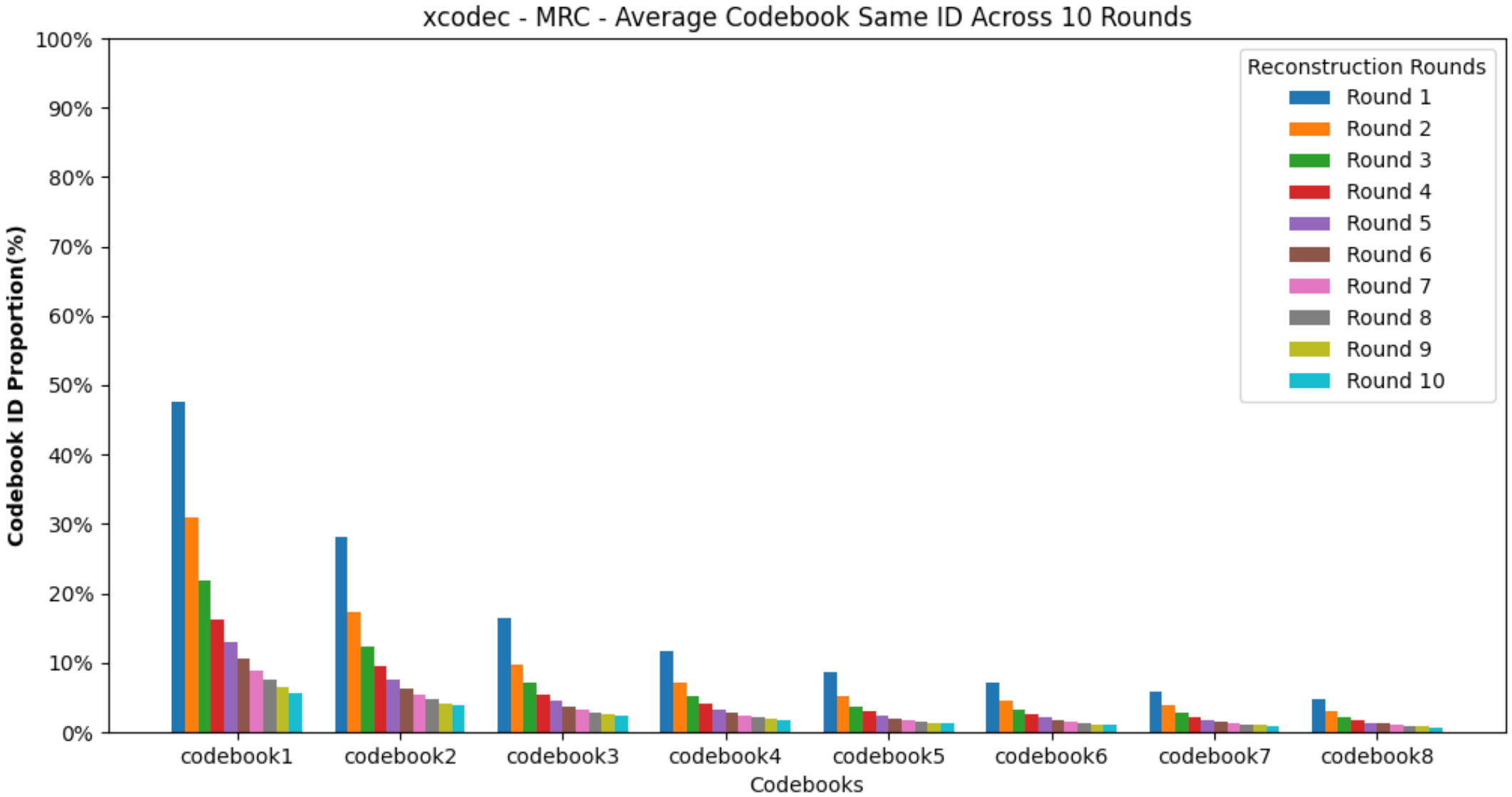} 
\caption{Multi-round Reconstruction results of XCodec.}
\label{fig6}
\end{figure}

\vspace{-16pt}

\begin{figure}[H]
\centering
\includegraphics[width=1.0\columnwidth]{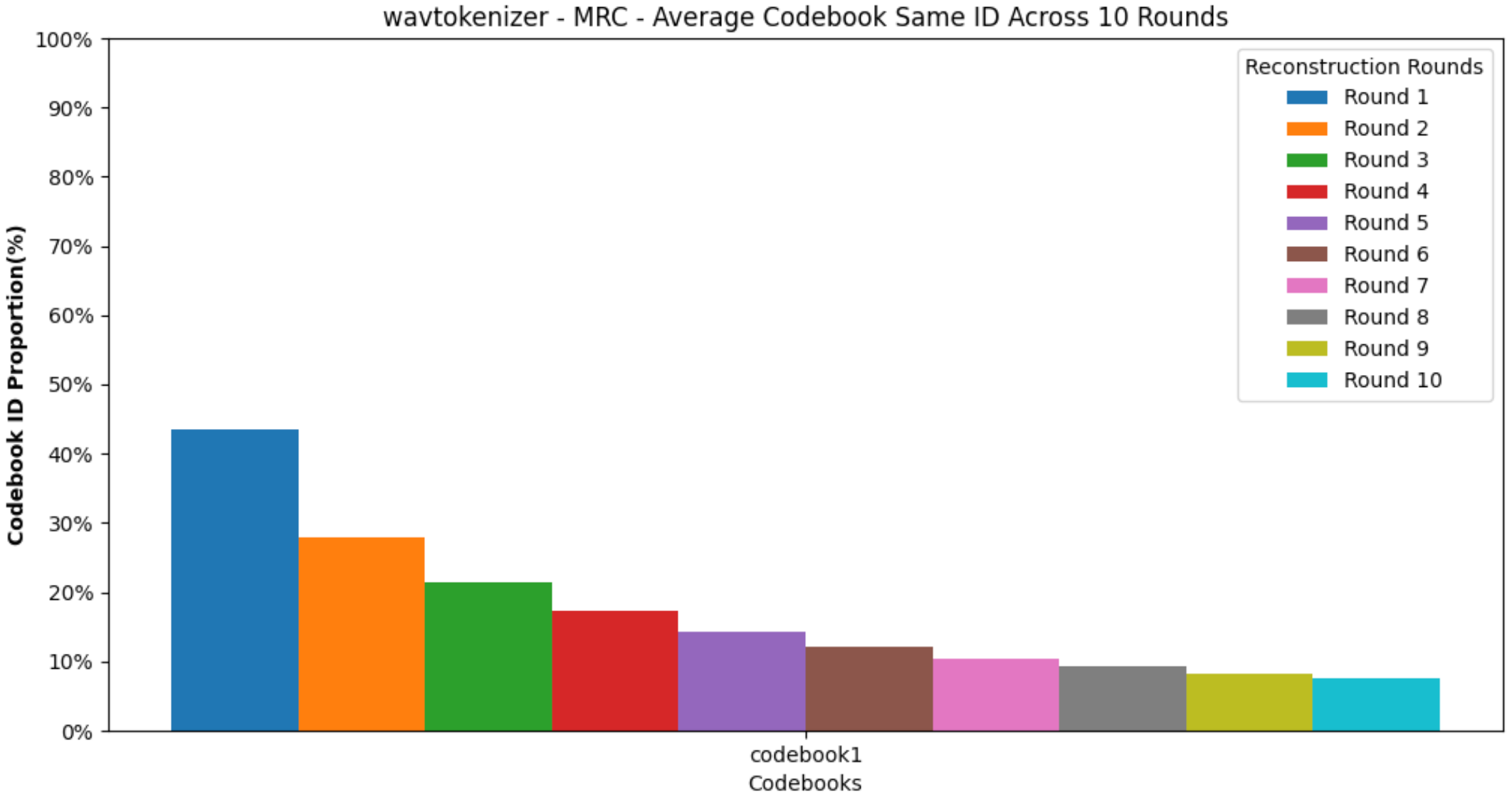} 
\caption{Multi-round Reconstruction results of WavTokenizer.}
\label{fig7}
\end{figure}

\vspace{-17pt}

\begin{figure}[H]
\centering
\includegraphics[width=1.0\columnwidth]{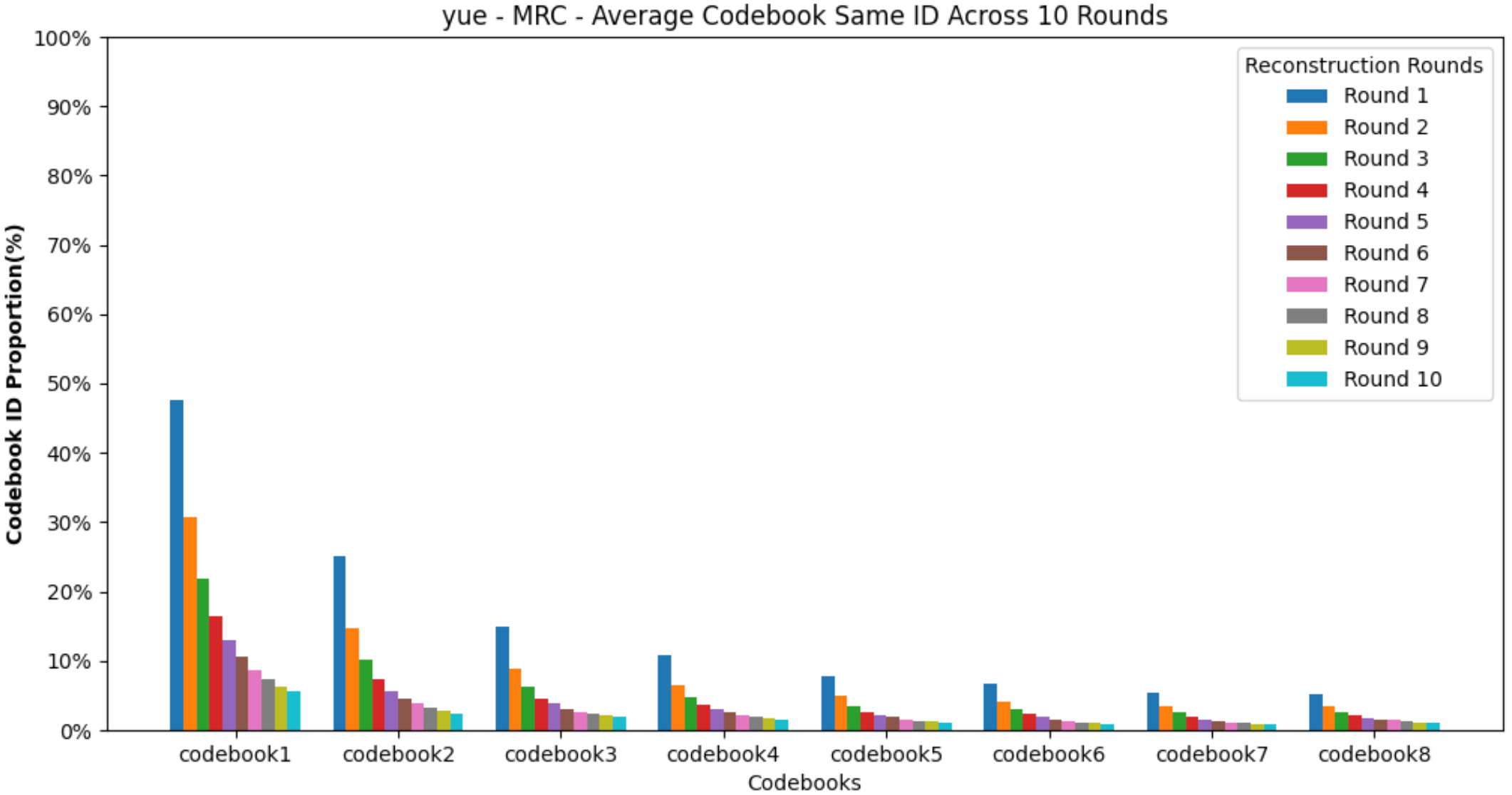} 
\caption{Multi-round Reconstruction results of YuE.}
\label{fig8}
\end{figure}

\section{Appendix B: Audio Time Shift results of different codecs}\label{Appendix_B}

\begin{figure}[H]
\centering
\includegraphics[width=1.0\columnwidth]{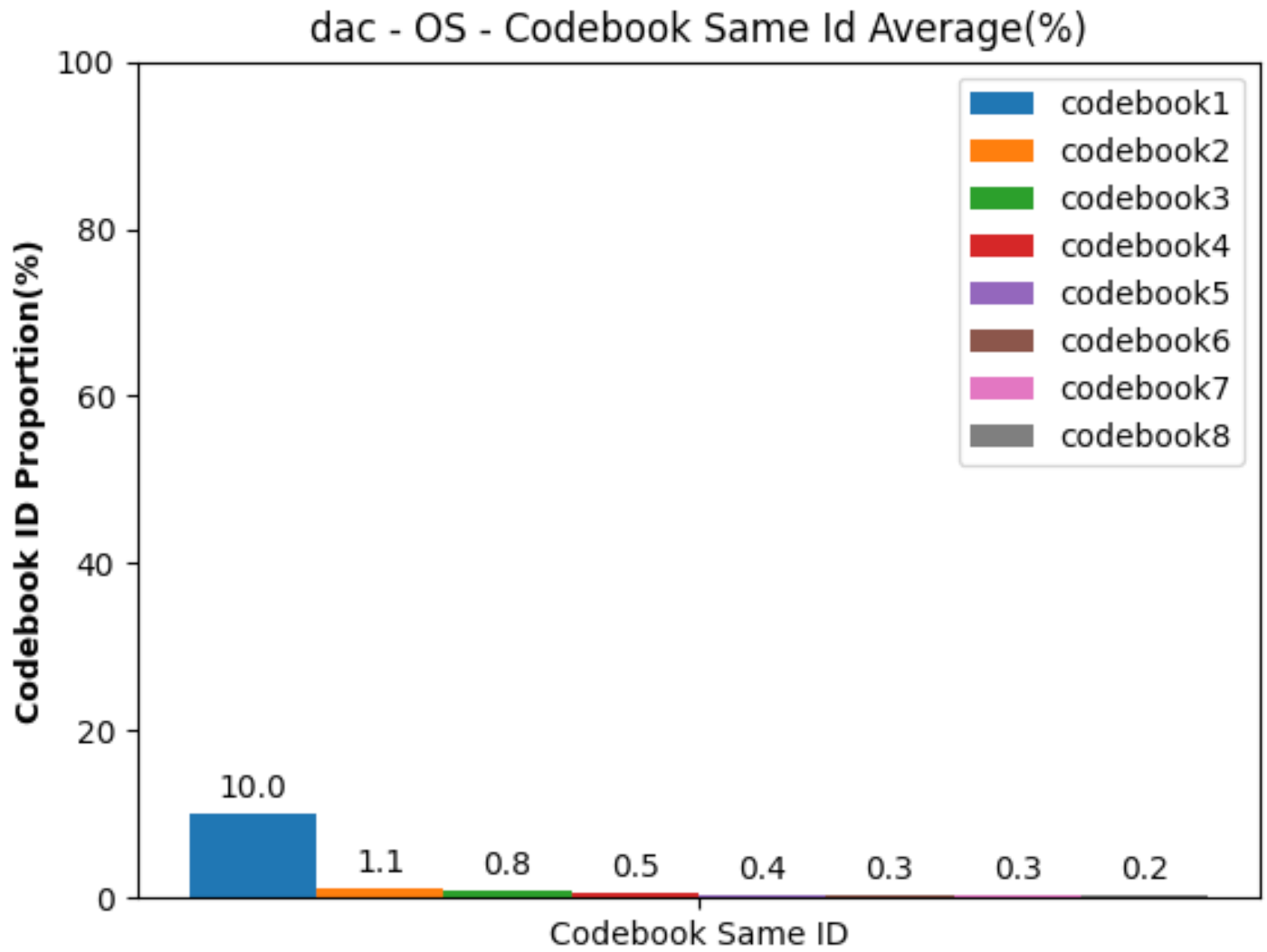} 
\caption{Audio Time Shift results of DAC.}
\label{fig9}
\end{figure}

\vspace{-16pt}

\begin{figure}[H]
\centering
\includegraphics[width=1.0\columnwidth]{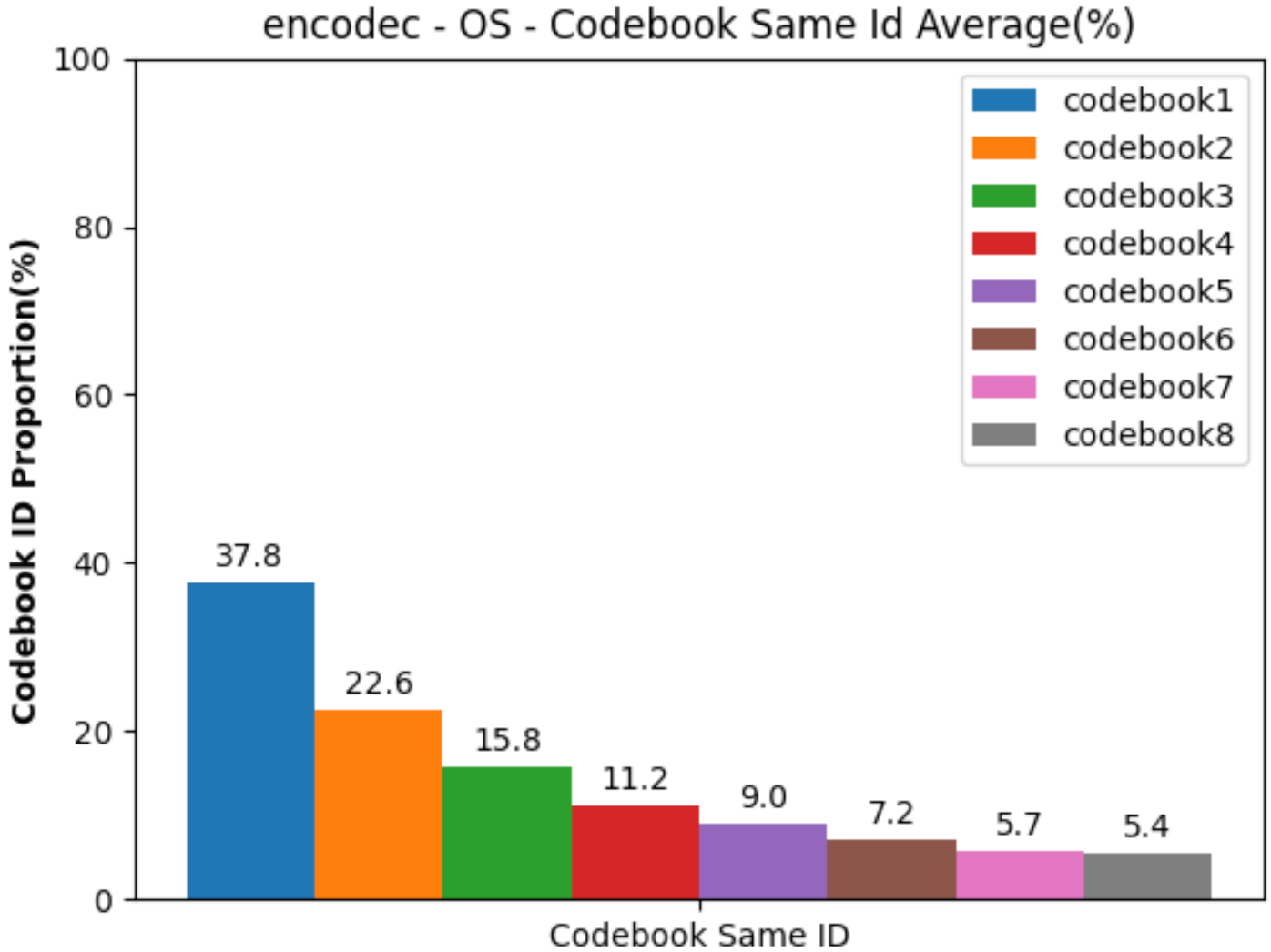} 
\caption{Audio Time Shift results of EnCodec.}
\label{fig10}
\end{figure}

\vspace{-16pt}

\begin{figure}[H]
\centering
\includegraphics[width=1.0\columnwidth]{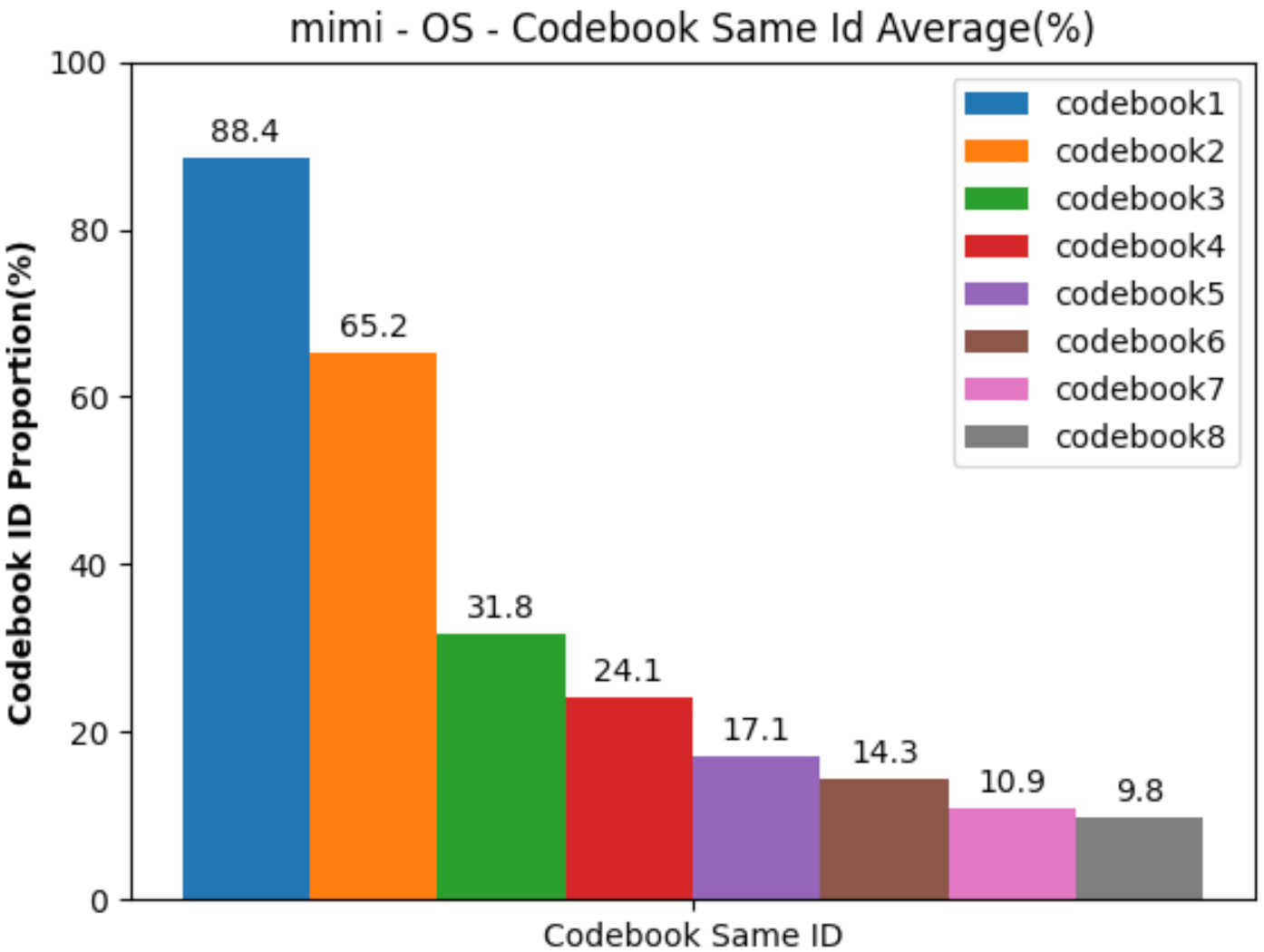} 
\caption{Audio Time Shift results of Mimi.}
\label{fig11}
\end{figure}

\vspace{-14pt}

\begin{figure}[H]
\centering
\includegraphics[width=1.0\columnwidth]{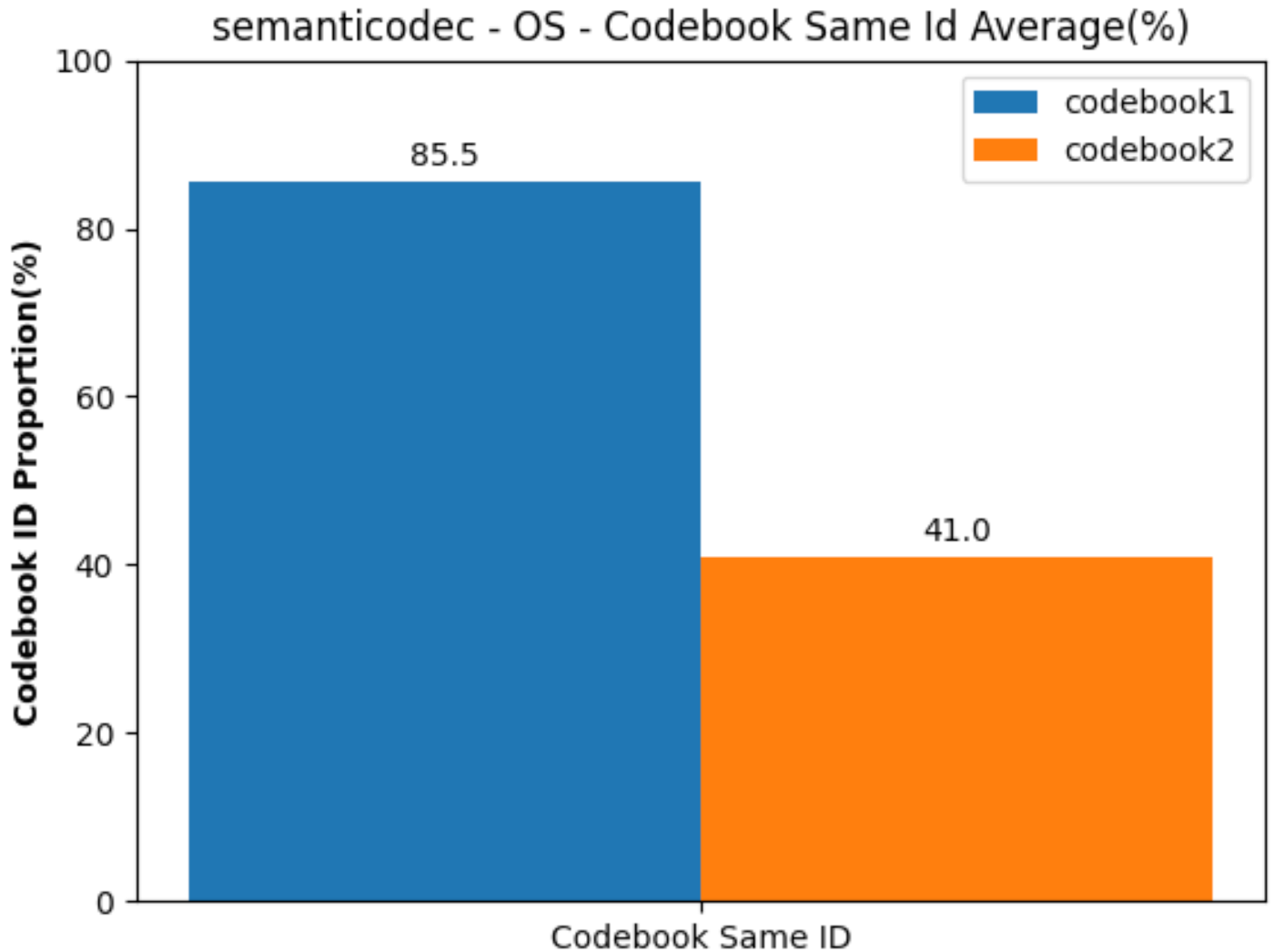} 
\caption{Audio Time Shift results of SemantiCodec.}
\label{fig12}
\end{figure}

\vspace{-14pt}

\begin{figure}[H]
\centering
\includegraphics[width=1.0\columnwidth]{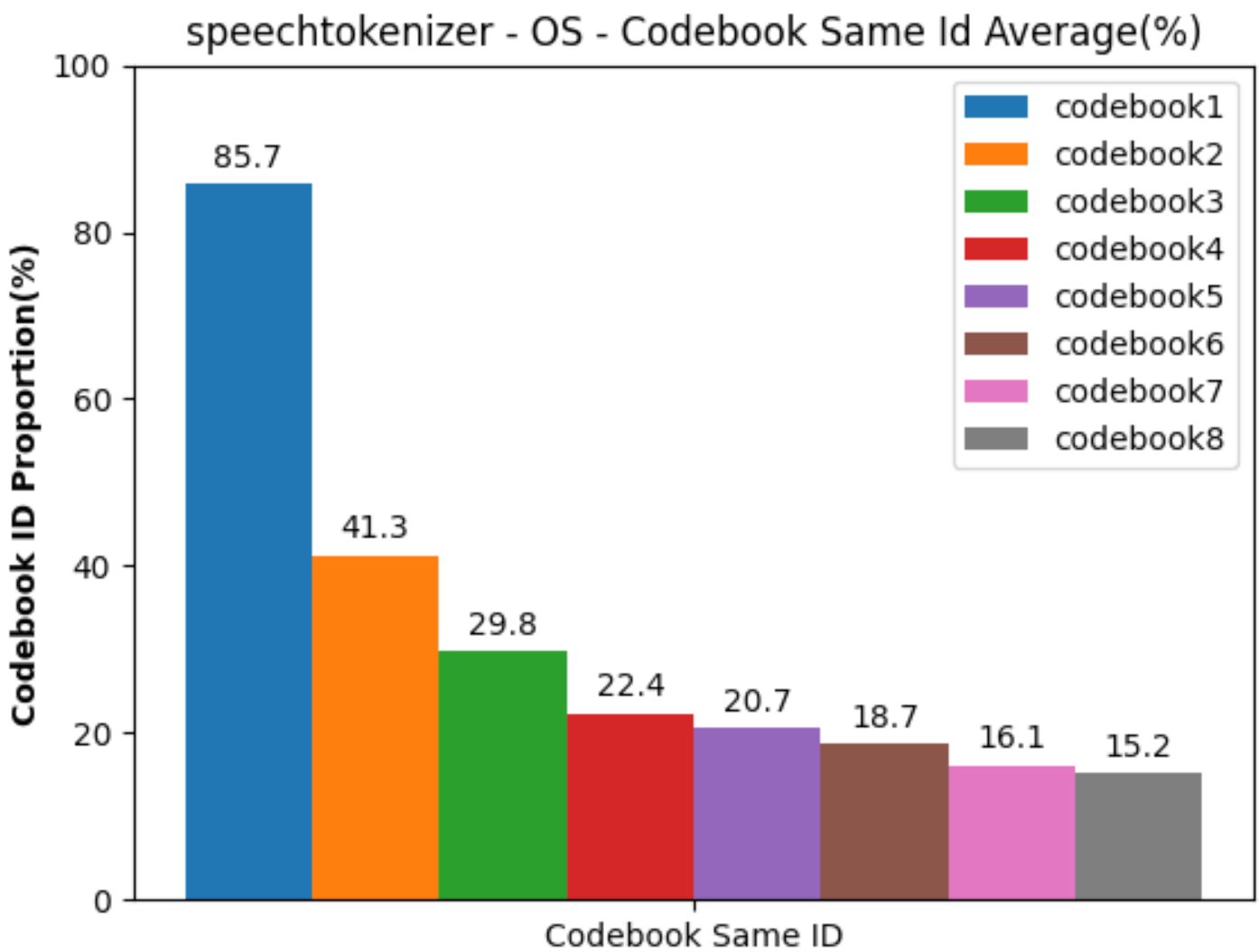} 
\caption{Audio Time Shift results of SpeechTokenizer.}
\label{fig13}
\end{figure}

\vspace{-14pt}

\begin{figure}[H]
\centering
\includegraphics[width=1.0\columnwidth]{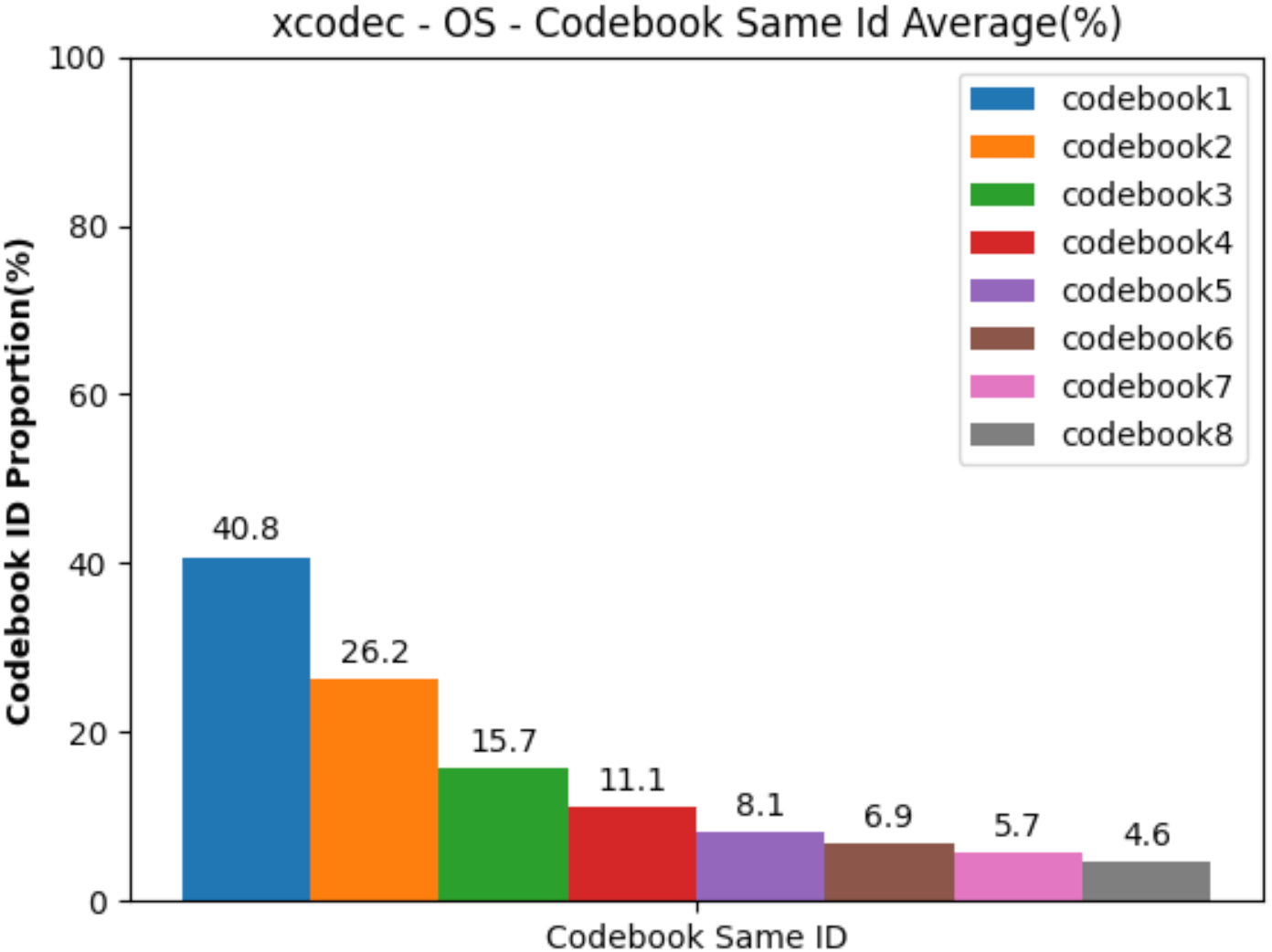} 
\caption{Audio Time Shift results of XCodec.}
\label{fig14}
\end{figure}

\vspace{-16pt}

\begin{figure}[H]
\centering
\includegraphics[width=1.0\columnwidth]{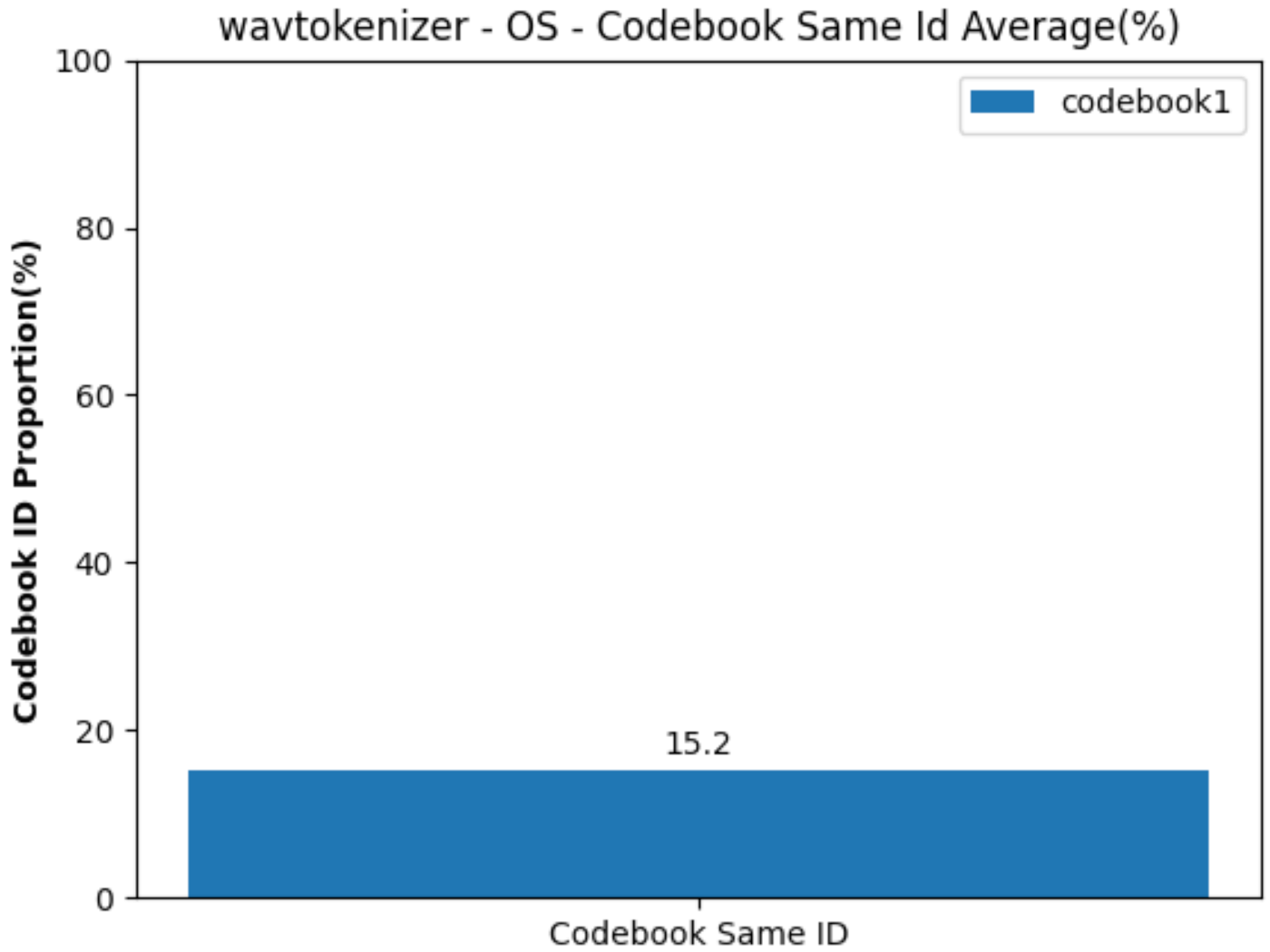} 
\caption{Audio Time Shift results of WavTokenizer.}
\label{fig15}
\end{figure}

\vspace{-10pt}

\begin{figure}[H]
\centering
\includegraphics[width=1.0\columnwidth]{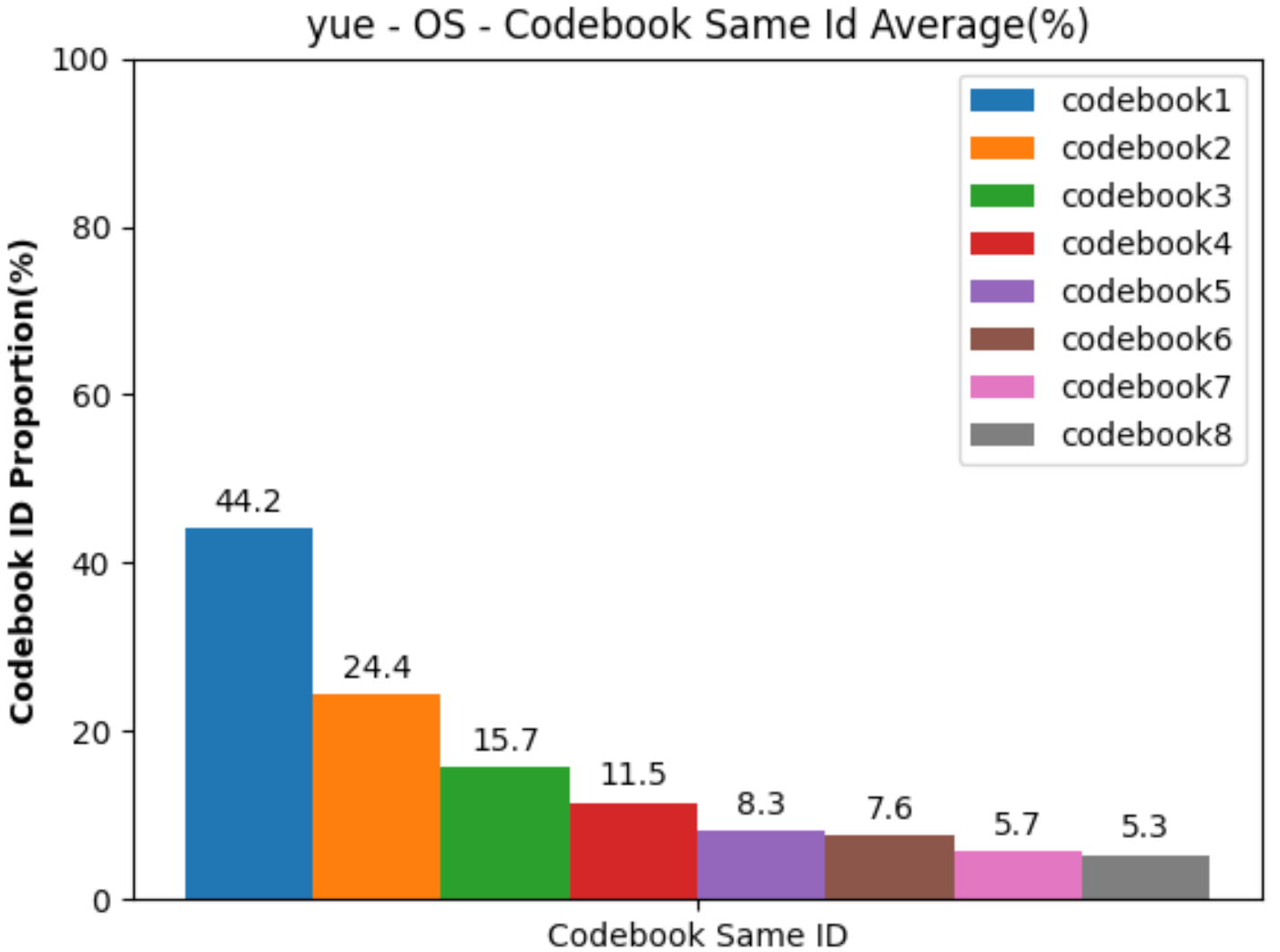} 
\caption{Audio Time Shift results of YuE.}
\label{fig16}
\end{figure}

\section{Appendix C: Visualization of music, speech and sound probe task results}\label{Appendix_C}

\begin{figure}[H]
\centering
\includegraphics[width=1\columnwidth]{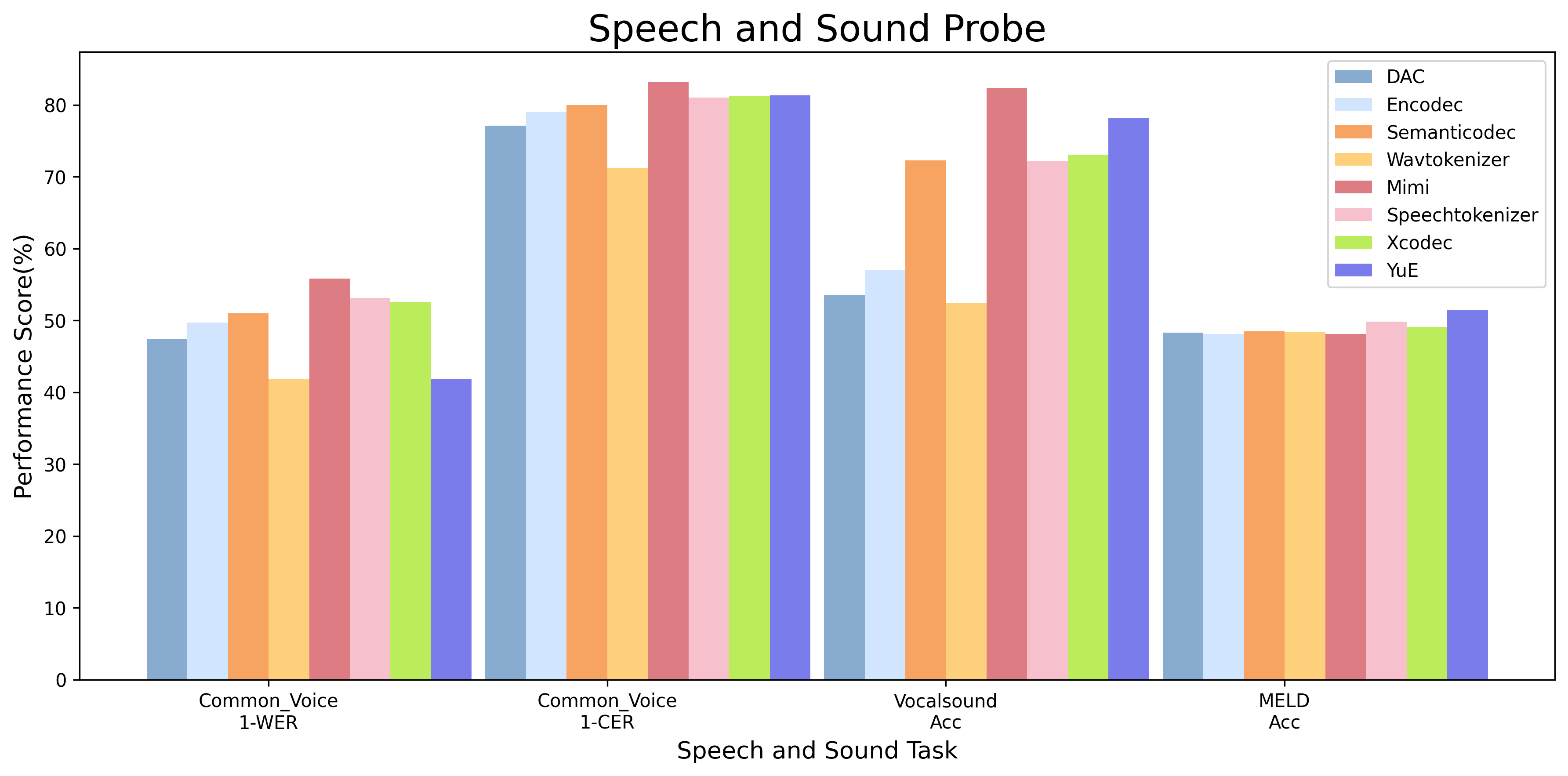} 
\caption{Visualization for the speech and sound probe tasks.}
\label{fig17}
\end{figure}

\begin{figure}[H]
\centering
\includegraphics[width=1\columnwidth]{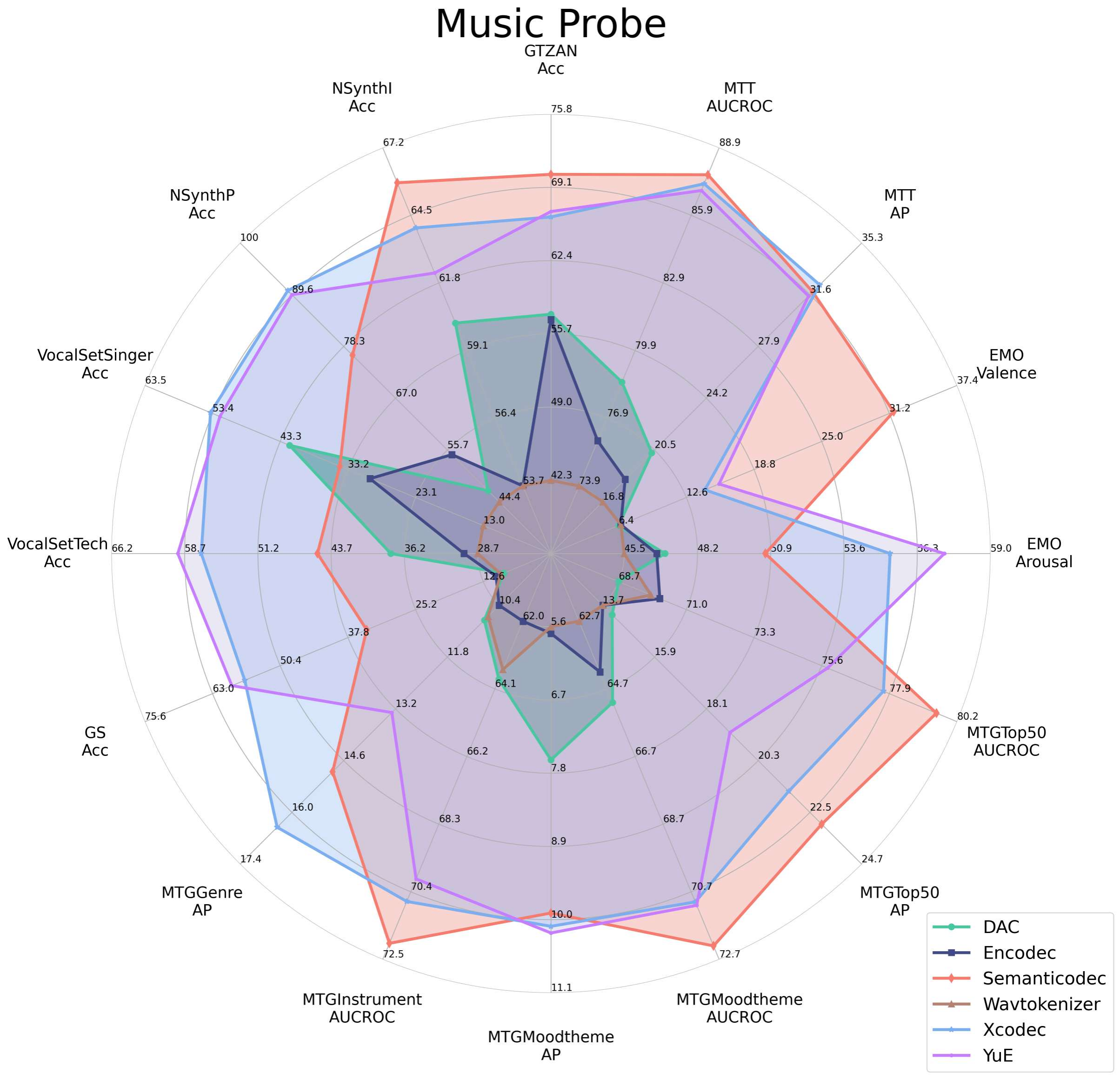} 
\caption{Visualization for the music probe tasks.}
\label{music_probe}
\end{figure}

\section{Appendix D: Introduction to downstream probe tasks and related datasets}\label{Appendix_D}
We integrate a comprehensive dataset consisting of 17 sub-datasets from 12 audio collections (mostly derived from the MARBLE benchmark), covering major audio categories of speech, environmental sound, and music. Based on this dataset, we conduct 11 different types of probe tasks to examine the performance of different codec representations across different audio information dimensions, such as emotion, linguistic content, acoustic scene, and speaker identity.

\textbf{Genre Classification (GC)}: This task aims to classify music audio into predefined genres (e.g., rock, pop, classical). We use the GTZAN dataset and adopt Accuracy (Acc) as the performance metric. Additionally, we utilize MTG-Genre, a subset of MTG-Jamendo. Considering its longer track durations, we take the first 150 seconds of each track, segment them into 10-second clips, and stack them to serve as the input for the codec. This approach balances computational resources with the evaluation requirements. We use the Area Under the ROC Curve (ROC-AUC) and Average Precision (AP) to evaluate the representation's ability to encode genre information.

\textbf{Key Detection (KD)}: The goal of key detection is to predict the musical key of a piece of music, which is defined by its pitch center and mode (e.g., C major, a minor). We use the GiantSteps Key dataset, a collection of electronic dance music containing 24 major and minor keys. We consider the musical key as a global feature of the audio, processing it by stacking 10-second segments as the codec's input. We then use Acc as the evaluation metric to assess the model's ability to capture information about the musical structure.

\textbf{Emotion Detection (ED)}: This task focuses on identifying the emotional state or dimension conveyed by the audio (e.g., happiness, sadness, anger). We integrate several datasets for this purpose: for the Valence and Arousal labels provided by the EmoMusic dataset, we employ a regression model for prediction and use the R² metric for evaluation. This helps assess the semantic information (high Valence) and acoustic information (high Arousal) embedded in the codec features. For the MTG MoodTheme dataset, which is a multi-label classification task with 59 emotion categories, we use ROC-AUC to evaluate the representation's ability to encode complex musical emotion information. Finally, using the MELD conversational speech dataset, we test the codec's capability to distinguish among seven basic emotions in a realistic context, which is evaluated with Acc.

\textbf{Vocal Technique Detection (VTD)}: This task aims to identify specific vocal techniques used by singers in musical compositions. It is a relatively uncommon, fine-grained identification task that focuses on the performance-level details. The main publicly available dataset is VocalSet, which contains recordings of 17 different vocal techniques performed by 20 professional singers, with each audio segment representing one technique category. We use Acc as the metric to evaluate the codec's ability to distinguish these subtle acoustic features.

\textbf{Pitch Classification (PC)}: This task aims to classify the main pitch content of a musical audio clip, with the range corresponding to MIDI note numbers 0 to 127 on the chromatic scale. We use the NSynth dataset, which consists of a large number of 4-second monophonic recordings. Due to its monophonic nature, this task can be viewed as a 128-class fine-grained pitch classification problem. It is designed to evaluate the accuracy of the codec's representation of fundamental frequency information, with performance assessed using Acc.

\textbf{Music Tagging (MT)}: This is a comprehensive evaluation task in the music domain that requires the model to assign multiple descriptive tags to music clips. These tags may cover various types, such as genre, instrument, and mood. We use the MagnaTagATune and MTG Top50 datasets. Following the MARBLE processing principles, we focus on our evaluation the model's ability to predict the Top 50 most frequent tags within these datasets. Given its multi-label nature, the final performance is measured by the ROC-AUC and the PR-AUC/AP. These metrics are used to evaluate the overall capability of the features in representing musical information.

\textbf{Instrument Classification (IC)}: This task aims to identify one or more musical instruments present in an audio recording. In the MARBLE classification system, this is considered an Acoustic-Level task, and its results evaluate the codec's ability to represent fundamental acoustic features. For evaluation, we use the NSynth dataset, which contains 11 single-instrument categories and is evaluated using Acc. We also use the MTG Instrument dataset, a multi-label collection with 41 labels, which is evaluated using ROC-AUC and PR-AUC/AP.

\textbf{Automatic Speech Recognition (ASR)}: This task focuses on transforming speech signals from audio recordings into textual content. We use the Common Voice dataset, which contains approximately 26119 hours of recordings, including a variety of demographic metadata such as age, gender, and accent. Among these, about 17127 hours of validated data cover 104 languages, with each language providing the necessary training, development, and test sets required to build a speech recognition model. Word Error Rate (WER) and Character Error Rate (CER) are used as the evaluation metrics. 

\textbf{Singer Identification (SI)}: This task aims to identify the singer's identity from a short music recording. For this task, we use the VocalSet dataset, a collection containing audio from 20 different singers. We follow the MARBLE-recommended dataset partition (training:validation:test = 12:8:5), and ensure that all singer categories are evenly distributed. Finally, Acc is used to evaluate the codec's ability to distinguish individual vocal features.

\textbf{Vocals Sound Classification (VSC)}: This task aims to classify various non-linguistic sounds made by humans. We use the VocalSound dataset, which contains six common non-speech human sounds: laughter, sighs, coughs, throat clearing, sneezes, and sniffs. Since the audio clips in the dataset have non-uniform lengths, we pad all audio to a uniform length before inputting them into the codec. The evaluation for this task is conducted using Acc.

\textbf{Environmental Sound Classification (ESC)}: This task focuses on identifying sounds from the environment. We use the ESC-50 dataset, which is a labeled collection of 2000 environmental audio recordings consisting of 5-second-long recordings divided into 50 semantic categories. Since the original dataset does not provide an official standard split, we use a 9:1 ratio to self-partition it into a training set and a test set, with Acc as the metric for the evaluation of this dataset.

\end{document}